\documentclass[12pt]{iopart} 

\def\>{\rangle}
\def\<{\langle}
\def\ave#1{\left\langle #1\right\rangle}
\def\aave#1{\<\!\< #1 \>\!\>}
\def\bbar#1{\bar{\bar{#1}}}
\def\d{{\rm d}}
\def\ii{{\rm i}}

\def\Z{{\mathbbm{Z}}}

\def\one{{\mathbbm{1}}}
\def\P{{\rm P}}

\def\S{{\rm S}}

\newcommand{\op}[1]{\hat{#1}}

\newcommand{\bra}[1]{\langle #1|}
\newcommand{\ket}[1]{|#1\rangle}
\newcommand{\braket}[2]{\langle #1|#2\rangle}

\def\vec#1{{\bm{#1}}}

\usepackage{bbm}
\usepackage{bm}
\usepackage{graphicx}
\usepackage{hyperref}

\begin{document}

\title[Quantum freeze of fidelity decay]
{Quantum freeze of fidelity decay for a class of integrable dynamics}

\author{Toma\v z Prosen and Marko \v Znidari\v c}

\address{Physics Department, Faculty of Mathematics and Physics,
University of Ljubljana, Ljubljana, Slovenia}

\eads{\mailto{prosen@fiz.uni-lj.si}, \mailto{znidaricm@fiz.uni-lj.si}}  

\begin{abstract}
We discuss quantum fidelity decay of classically regular dynamics, in particular for an 
important special case of a vanishing time averaged perturbation operator, {\em i.e.}
vanishing expectation values of the perturbation in the eigenbasis of
unperturbed dynamics. A complete semiclassical picture of this situation 
is derived in which we show that the quantum fidelity of {\em individual coherent 
initial states} exhibits three different regimes in time:
(i) first it follows the corresponding classical fidelity up to time 
$t_1 \sim \hbar^{-1/2}$, (ii) then it freezes on a plateau of 
constant value, (iii) and after a time scale 
$t_2 \sim \min\{\hbar^{1/2}\delta^{-2},\hbar^{-1/2}\delta^{-1}\}$ 
it exhibits fast ballistic decay as 
$\exp(-{\rm const}\,\delta^4 t^2/\hbar)$
where $\delta$ is a strength of perturbation.
All the constants are computed in terms of classical dynamics for 
sufficiently small effective value $\hbar$ of the Planck constant.
A similar picture is worked out 
also for general initial states, and specifically for random initial states,
where $t_1 \sim 1$, and $t_2 \sim \delta^{-1}$. This prolonged stability of quantum dynamics 
in the case of a vanishing time averaged perturbation could prove to be useful in designing 
quantum devices. Theoretical results are verified by numerical experiments on 
the quantized integrable top.
\end{abstract}

\submitto{\NJP}
\pacs{03.65.Yz, 03.65.Sq, 05.45.Mt}

\section{Introduction}

Squared modulus of the overlap between a pair of time evolving quantum states propagated by two slightly
different Hamiltonians, known as the fidelity or the quantum Loschmidt echo,
has recently attracted a lot of attention \cite{peres,jalabert,generalrefs,Prosen01,QC,PZ,golden}. In addition to numerous 
numerical simulations, several theoretical results have been proposed to describe the 
fidelity decay in relation to the nature of the (corresponding classical) dynamics.
Jalabert and Pastawski \cite{jalabert} have related fidelity decay for coherent initial states
at very short times, namely below or around the Ehrenfest time $\propto\log\hbar$, 
to the classical phase space stretching rate as characterized by the Lyapunov 
exponents. 
In more general situations, and in particular for longer time scales,
fidelity decay has been related to the integrated time correlation function
of the perturbation through a kind of fluctuation-dissipation relationship
\cite{Prosen01,QC,PZ}.

In a recent paper \cite{PZ} we have developed a general theory of 
fidelity decay based on a semiclassical treatment of this fluctuation-dissipation formula.
It turns out that if the corresponding classical dynamics is fully chaotic then
the decay of fidelity is, after a short $\sim \log\hbar$ (Ehrenfest) timescale, 
independent of the structure of the initial state in accordance with the quantum ergodicity. 
On the contrary, if the corresponding classical dynamics is
regular then the long time asymptotics sensitively depend on the structure
of the initial state and range from a {\em Gaussian} fidelity decay for 
{\em coherent initial states} 
to a {\em power-law} fidelity decay for {\em random initial states}.
For regular classical dynamics the theory \cite{PZ} predicts {\em faster}
decay of fidelity on a short time scale $\propto \delta^{-1}$ ($\delta=$strength of the
perturbation), as the time correlation function of the perturbation observable 
does not decay, {\em compared} to the chaotic classical dynamics where the decay time scale is 
longer  $\propto\delta^{-2}$, and is longer the faster the decay of correlations we have.
However, the fast --- ballistic decay of fidelity in the case of regular classical dynamics described by the theory \cite{PZ} does not happen in one special but important case, namely when the time average of the perturbation ({\em i.e.} the observable which 
perturbs the Hamiltonian) vanishes. Classically, this means that the perturbation does not change the 
frequencies of the invariant (KAM) tori in the leading order in $\delta$, at least in the phase space region of interest.

Such a case of regular classical dynamics with vanishing time-averaged perturbation is the subject of the present paper. Though this is not a generic case for a sufficiently
large class of perturbations, it may emerge naturally if the system and the perturbation
possess appropriate discrete or continuous symmetries.
We will discuss general initial states, and specifically also coherent and random initial states.
We find a very surprising result, namely that the quantum fidelity, after decaying for a short time (e.g. following
the classical fidelity \cite{PZ,Veble,Bruno} for coherent initial states), {\em freezes} on a plateau of constant value.
This is purely a quantum effect and has no analogue in the classical fidelity.
The relative time span of the plateau is of the order of inverse perturbation strength $1/\delta$ and
can be made arbitrary large for small perturbations. However for long times after the plateau ends, 
the fidelity displays a ballistic decay with the characteristic time scale $\propto 1/\delta^2$, {\em e.g.} 
Gaussian for coherent initial states and power law $t^{-d}$ for random initial states where
$d$ is the number of freedoms. This ballistic decay can be explained
semiclassically due to perturbative changes of the frequencies of invariant tori in the second order in 
$\delta$.
For coherent initial states in one dimension we find and explain another quite surprising general
phenomenon, which we call `the echo resonance' where the fidelity displays sudden and significant revivals which can, 
under certain conditions, come back even to value $1$.
This happens at particular values of times, which depend on the derivative of the classical frequency with respect to the 
canonical action and do not depend on $\delta$.

Using the formalism of action-angle variables and its semiclassical quantization we derive
explicit semiclassical formulae, in the leading order in $\hbar$, 
for the fidelity in all the regimes. Our results are demonstrated with high precision 
by numerical experiments using a regular quantum top which is perturbed by 
periodic kicking.
The quantum saturation of fidelity, which is a central result of this paper, 
may also be of some practical importance as it provides a mechanism for stabilizing the regular 
quantum dynamics.

In section 2 we define the basic quantities and
study the general properties of the so-called {\em echo
operator} whose expectation value gives the fidelity.
We propose a useful asymptotic ansatz for the echo operator, which is used later 
in section 3, in combination with the semiclassical action-angle dynamics, to derive 
explicit general results on the echo operator and fidelity and to identify different regimes.
In section 4 we define a numerical model on which the results for two specific classes of 
initial states, namely coherent and random states, are later quantitatively validated in sections 5 
and 6, respectively. In section 7 we discuss the general picture and summarize the results.

\section{Quantum mechanics of the echo operator}

Let $H_0$ and $H_\delta = H_0 + \delta\cdot H'$ denote the {\em unperturbed} and the 
{\em perturbed} Hamiltonian, respectively. In order to cover the even more general case 
of periodically time-dependent (e.g. kicked) systems, say of period $\tau$,
$H_\delta(\tau'+\tau)=H_\delta(\tau')$, we utilize our formalism in terms of the Floquet map 
$U_\delta=\left[\op{\cal T}\exp(\ii \int_0^\tau\d\tau' H_\delta(\tau')/\hbar)\right]^\dagger$,
where $\op{\cal T}$ denotes {\em left-to-right} time ordered product.
Dynamics is now generated by a discrete group $U_\delta^t, t\in\Z$, where an autonomous
continuous time flow is approached in the limit $\tau \to 0$.
It seems convenient to postulate slightly different but completely
general form of a {\em small} perturbation
\begin{equation}
U_\delta = U_0 \exp(-\ii V \tau \delta/\hbar)
\end{equation}
generated by a hermitean operator $V$ which in the leading order 
matches $H'$, $V = H' + {\cal O}(\tau\delta)$.
We note that all the results in the paper can be trivially translated to the continuous 
time case by substituting $t\tau \to T$, $\tau\sum_{t'=0}^{t-1} \rightarrow \int_0^T\d\tau'$.

Starting from the same initial state $\ket{\psi}$, the {\em fidelity} or the {\em Loschmidt 
echo $F(t)$} is defined as the squared modulus of the overlap between 
$U_0^t\ket{\psi}$ and $U_\delta^t\ket{\psi}$, 
namely
\begin{equation}
F(t) = |f(t)|^2, \qquad f(t) = \bra{\psi}M_\delta(t)\ket{\psi},
\end{equation}
where $f(t)$ is called the {\em fidelity amplitude} and
\begin{equation}
M_\delta(t)=U_\delta^{-t}U_0^t
\end{equation}
is the {\em echo operator}. Equivalently, $M_\delta(t)$ is the time-ordered 
propagator generated by the perturbation $V_t = U_0^{-t} V U_0^t$ in the interaction picture
\cite{Prosen01,PZ}
\begin{equation}
M_\delta(t) = \prod_{t'=0}^{t-1}{\exp{\left( \ii\frac{\tau\delta}{\hbar} V_{t'} \right)}} =\hat{\cal T} \exp\left(\ii\frac{\tau\delta}{\hbar}
\sum_{t'=0}^{t-1} V_{t'} \right).
\label{eq:echoop}
\end{equation}
The essential results on the behaviour of fidelity \cite{PZ} are then derived from 
the combination
of perturbative and semiclassical considerations of the formula (\ref{eq:echoop}).
For example, the essential physics is contained in a linear response approximation
which is obtained by expanding (\ref{eq:echoop}) to second order in $\delta$,
\begin{eqnarray}
F(t) &=& 1 - \frac{\tau^2\delta^2}{\hbar^2}\sum_{t',t''=0}^{t-1}C(t',t'')+\ldots,
\label{eq:LR}\\
C(t',t'') &=& \ave{V_{t'}V_{t''}} - 
\ave{V_{t'}}\ave{V_{t''}},
\label{eq:Ct}
\end{eqnarray}
where $\ave{\bullet}:=\bra{\psi}\bullet\ket{\psi}$ is the expectation value in the
initial state $\ket{\psi}$. Thus stronger decay of correlations qualitatively enhances the
stability of quantum motion \cite{Prosen01,QC,PZ}. It is useful to rewrite the 
double-sum on the RHS of linear response formula (\ref{eq:LR}) in terms of the 
uncertainty of the {\em integrated perturbation operator}
\begin{equation}
\Sigma_t = \tau\sum_{t'=0}^{t-1} V_{t'},
\label{eq:sigdef}
\end{equation}
namely
\begin{equation}
F(t) = 1 - \frac{\delta^2}{\hbar^2}\left\{\ave{\Sigma^2_t}-\ave{\Sigma_t}^2\right\} 
+ {\cal O}(\delta^4).
\label{eq:LR2}
\end{equation}

Here we take a slightly different route and apply the Baker-Campbell-Hausdorff (BCH) 
expansion $e^A e^B = \exp(A + B + (1/2)[A,B] + \ldots)$
to the echo operator (\ref{eq:echoop})
\begin{eqnarray}
M_\delta(t) &=& \exp\left\{\ii\frac{\tau\delta}{\hbar}\sum_{t'=0}^{t-1} V_{t'} - 
\frac{\tau^2\delta^2}{2\hbar^2}\sum_{t'=0}^{t-1}\sum_{t''=t'}^{t-1} [V_{t'},V_{t''}] 
+ \ldots \right\}
\nonumber\\
&=& \exp\left\{\frac{\ii}{\hbar}\left(
\Sigma_t \delta + \frac{1}{2}\Gamma_t\delta^2 + \ldots\right)
\right\}
\label{eq:BCH}
\end{eqnarray}
where $[A,B]:=AB-BA$, introducing another operator valued series
\begin{equation}
\Gamma_t = \frac{\ii\tau^2}{\hbar}
\sum_{t'=0}^{t-1}\sum_{t''=t'}^{t-1} [V_{t'},V_{t''}].
\label{eq:Gamma}
\end{equation}
Note that for systems with a well defined classical limit the operator 
$\Gamma_t$ corresponds to $\hbar$-independent classical observable as
$-\ii/\hbar [\bullet,\bullet]$ corresponds to the classical Poisson bracket.
In the ergodic and mixing case, of say classically strongly 
chaotic dynamics, straightforward expansion of the exponential (\ref{eq:echoop}) 
gives the 
Fermi-golden-rule \cite{golden} exponential decay 
$F(t) = \exp(-\kappa t)$ \cite{Prosen01,PZ}
where the argument $\kappa t$ is precisely the double-sum of
correlation function on RHS of (\ref{eq:LR}) for sufficiently long times.
However in the opposite case of classically regular (integrable) 
dynamics, on which we focus in the following sections of this paper, 
the BCH form (\ref{eq:BCH}) will turn out to be particularly useful.

Let us first generally discuss the expression (\ref{eq:BCH}) from the
point of view of exact unitary quantum dynamics. 
For a typical observable $V$, one can define a nontrivial time-average operator
\begin{equation}
\bar{V} = \lim_{t\to\infty}\frac{\Sigma_t}{t\tau} = 
\lim_{t\to\infty}\frac{1}{t}\sum_{t'=0}^{t-1} V_{t'},
\label{eq:Vave}
\end{equation}
which is by construction an invariant of motion, $[U_0,\bar{V}] = 0$.
In a generic case of {\em non-degenerate spectrum} of $U_0$, 
the time average is simply the diagonal part in the eigenbasis 
$\ket{n}$ of the unperturbed evolution,
$U_0\ket{n} = e^{-\ii\varphi_n}\ket{n}$, namely
\begin{equation}
\bar{V} = \sum_n V_{nn} \ket{n}\bra{n},
\end{equation}
where $V_{nm} := \bra{n}V\ket{m}$. In general we split the perturbation into a sum of diagonal and residual part
\begin{equation}
V = \bar{V} + V_{\rm res}.
\end{equation}
We say that the observable $V$ is {\em residual} if $V=V_{\rm res}$.
This corresponds to ergodicity of this specific observable, namely $\bar{V}=0$,
meaning that $V$ has zero diagonal elements, 
and this is clearly a special (non-generic) situation.\footnote[8]{We can always choose the perturbation $V$ to be
{\em traceless}, since subtracting a constant $V \to V - \one \tr V/\tr\one$ only changes the phase of the
amplitude $f$ and does not affect the fidelity $F = |f|^2$.}
In this paper we discuss integrable dynamics and the class of residual perturbations.
For non-degenerate eigenphases 
$\{\varphi_n\}$ the matrix elements 
of the second order term (\ref{eq:Gamma}) in BCH expansion 
can be straightforwardly calculated in the leading order in $t$ as
\begin{eqnarray}
\!\!\!\!\!\!\!\!\!\frac{\bra{n}\Gamma_t\ket{n}}{t\tau} &=& \frac{\tau}{\hbar} \sum_k^{k\neq n}
|V_{nk}|^2 \cot[{\textstyle\frac{1}{2}}(\varphi_k-\varphi_n)] + 
{\cal O}(t^{-1}),\label{eq:gammame}\\
\!\!\!\!\!\!\!\!\!\frac{\bra{n}\Gamma_t\ket{m}}{t\tau} &=& \frac{\tau}{\hbar}(V_{nn}\!-\!V_{mm})V_{nm}
\frac{e^{-\ii\frac{1}{2}(\varphi_n\!-\!\varphi_m)}\!+\!e^{-\ii(\varphi_n\!-\!\varphi_m)(\frac{1}{2}-t)}}
{2\sin[\frac{1}{2}(\varphi_m-\varphi_n)]} + {\cal O}(t^{-1}), \quad n\neq m.\nonumber
\end{eqnarray}
Hence we see that, provided the perturbation is residual, $\bar{V}=0$,
the limit of {\em doubly-averaged perturbation} defined as
\begin{equation}
\bbar{V} = 
\lim_{t\to\infty}\frac{\Gamma_t}{t\tau}
= \frac{\ii\tau}{\hbar}\lim_{t\to\infty}\frac{1}{t}\sum_{t'=0}^{t-1}\sum_{t''=t'}^{t-1} 
[V_{t'},V_{t''}]
\label{eq:defbbarV}
\end{equation}
exists and is {\em diagonal} in the eigenbasis of $U_0$:
\begin{equation}
\bbar{V} =\sum_n \bbar{V}_{nn} \ket{n}\bra{n},\qquad
\bbar{V}_{nn} = 
\frac{\tau}{\hbar}\sum_{k}^{k\neq n}
|V_{nk}|^2 \cot[{\textstyle\frac{1}{2}}(\varphi_k-\varphi_n)].
\label{eq:Vbar}
\end{equation}
Note that $\bbar{V}$ is again an invariant of motion, $[U_0,\bbar{V}]=0$, and
that, unlike for the time average $\bar{V}$, its trace always vanishes $\tr\bbar{V}=0$.

In the {\em generic} case, $\bar{V}\neq 0$ is a non-trivial operator.
For sufficiently small perturbation the second term in the exponent of RHS of 
(\ref{eq:BCH}) can always be neglected, since 
its arbitrary (finite) norm grows as $\propto t \delta^2$ following from 
$|\bra{n}\Gamma_t\ket{m}| < {\rm const}\times t$ [see eq. (\ref{eq:gammame})],
in comparison to the first term whose norm grows as 
$\propto t \delta$. 
For sufficiently long times, {\em i.e.} longer than the effective convergence time of the
limit (\ref{eq:Vave}), we can write $\Sigma_t \to t\tau \bar{V}$ so 
the echo operator can be written as $M_\delta(t) = \exp(\ii\bar{V}t\tau \delta/\hbar)$ 
from which useful
semiclassical expressions for different types of initial states 
were derived \cite{PZ}, all showing fidelity decay on an effective time scale 
$\propto \delta^{-1}$.
In the specific case of {\em residual} perturbation, $\bar{V} = 0$, the norm of
the first term in the exponential on RHS of (\ref{eq:BCH}) does not grow in time, 
as we shall discuss in the next section, so the second term will dominate for 
sufficiently long times. 

Although residual perturbations are not generic in the entire set of 
physically admissible perturbations $V$, they may nevertheless be of particular 
interest in cases where one is allowed to shift the entire diagonal part of the 
matrix $V_{nm}$ to the unperturbed Hamiltonian matrix, which is diagonal by definition. 
Also, it is easy to imagine practically or experimentally important situations
where vanishing of the diagonal part, $V_{nn} \equiv 0$, is required by 
the {\em symmetry}. For example, it is obvious that having a unitary symmetry operation $R$, 
$R^\dagger R=1$, commuting with the unperturbed evolution, $[R,U_0]=0$, and the perturbation $V$ which has
a negative 'parity' with respect to the symmetry operation,
$R^\dagger V R = -V$, is sufficient to give $V_{nn} = 0$.

As the case of generic perturbations has been treated in detail in 
previous publications \cite{Prosen01,PZ}, we shall from now on entirely 
concentrate on the residual case $\bar{V}=0$, 
unless explicitly stated otherwise. In this case we have
thus found the following uniform approximation of the echo operator
\begin{equation}
M_\delta(t) = \exp\left\{\frac{\ii}{\hbar}\left(\Sigma_t\delta + 
\frac{1}{2}\bbar{V}t\tau\delta^2\right)
\right\},
\label{eq:BCH1}
\end{equation}
which is accurate, for sufficiently small $\delta$, up to long times at least
of the order $1/\delta^2$. 
This is a consequence of the fact, that for $\bar{V}=0$ the third-order 
term in BCH expansion (\ref{eq:BCH})
again grows only linearly in time $\sim t \delta^3$,
and that the fourth-order term cannot grow faster than $\sim t^2\delta^4$.
The rest of the paper will be dedicated to the semiclassical exploration of the formula 
(\ref{eq:BCH1}).
 
\section{Semiclassical asymptotics}

Since the classical mechanics is assumed to be completely integrable
(at least locally, by KAM theorem, in the phase space part of interest) we
can write the classical limit $v(\vec{j},\vec{\theta})$ 
of the perturbation operator $V$ in canonical action-angle variables 
$\{j_k,\theta_k,k=1\ldots d\}$
in $d$ degrees of freedom as the Fourier series in $d$ dimensions
\begin{equation}
v(\vec{j},\vec{\theta}) = \sum_{\vec{m}\in\Z^d} v_{\vec{m}}(\vec{j}) 
e^{\ii\vec{m}\cdot\vec{\theta}}.
\label{eq:fourier}
\end{equation}
We shall throughout the paper use lower/upper case letters to denote the
corresponding classical/quantum observables.
Note that the classical limit of the unperturbed Hamiltonian $H_0$ can be written as
a function $h_0(\vec{j})$ of the canonical actions only, yielding the well-known quasi-periodic 
solution of the Hamilton's equations
\begin{eqnarray}
\vec{j}_t &=& \vec{j}, \nonumber\\ 
\vec{\theta}_t &=& \vec{\theta} + \vec{\omega}(\vec{j})t \pmod{2\pi}
\label{eq:hamileq}
\end{eqnarray}
with the dimensionless frequency vector
\begin{equation}
\vec{\omega}(\vec{j}) := \tau\frac{\partial h_0(\vec{j})}{\partial\vec{j}}.
\end{equation}
The classical limit of the time-averaged perturbation $\bar{V}$ is $\bar{v} = v_{\vec{0}}(\vec{j})$ which is
here assumed to vanish $v_{\vec{0}}(\vec{j}) \equiv 0$. 

In quantum mechanics, one quantizes the action-angle variables using the famous
EBK procedure \cite{EBK} where one defines the action (momentum) operators $\vec{J}$
and angle operators $\exp(\ii\vec{m}\cdot\vec{\Theta})$ satisfying the canonical
commutation relations
\begin{equation}
[J_k,\exp(\ii\vec{m}\cdot\vec{\Theta})] = \hbar m_k \exp(\ii\vec{m}\cdot\vec{\Theta}), \qquad k = 1,\ldots,d.
\label{eq:com}
\end{equation}
As the action operators are mutually commuting they have a common eigenbasis 
$\ket{\vec{n}}$ labeled by $d$-tuple of quantum numbers $\vec{n}=(n_1,\ldots,n_d)$,
\begin{equation}
\vec{J}\ket{\vec{n}} = \hbar (\vec{n} + \vec{\alpha})\ket{\vec{n}}
\end{equation}
where $0 \le \alpha_k \le 1$ are the Maslov indices 
%determined by the number of caustics in the projection of tori onto the configuration space.
which are irrelevant for the leading order semiclassical approximation employed in this paper.
It follows from eq. (\ref{eq:com}) that the angle operators act as shifts
\begin{equation}
\exp(\ii \vec{m}\cdot\vec{\Theta})\ket{\vec{n}} = \ket{\vec{n}+\vec{m}}.
\label{eq:shift}
\end{equation}
The Heisenberg equations of motion can be trivially solved, 
disregarding the operator ordering problem in the leading semiclassical order,
\begin{eqnarray}
\vec{J}_t &=& e^{\ii H_0 \tau t/\hbar}\vec{J} e^{-\ii H_0 \tau t/\hbar} = \vec{J},\nonumber\\
e^{\ii\vec{m}\cdot\vec{\Theta}_t} 
&=& e^{\ii H_0 \tau t/\hbar}e^{\ii\vec{m}\cdot\vec{\Theta}} e^{-\ii H_0 \tau t/\hbar} \cong 
e^{\ii\vec{m}\cdot\vec{\omega}(\vec{J})t} e^{\ii\vec{m}\cdot\vec{\Theta}},
\end{eqnarray}
in terms of the frequency operator $\vec{\omega}(\vec{J})$. In the whole paper we use the symbol $\cong$ for 
'semiclassically equal', {\em i.e.} asymptotically equal in the leading order in $\hbar$.
Similarly, time evolution of the perturbation observable is obtained in the 
leading order by substitution of classical with quantal action-angle 
variables in the expression (\ref{eq:fourier})
\begin{equation}
V_t = e^{\ii H_0 \tau t/\hbar}V e^{-\ii H_0 \tau t/\hbar} \cong
\sum_{\vec{m}\neq\vec{0}} v_{\vec{m}}(\vec{J}) 
e^{\ii\vec{m}\cdot\vec{\omega}(\vec{J})t}e^{\ii\vec{m}\cdot\vec{\Theta}}.
\label{eq:scVt}
\end{equation}

Now we are ready to write the semiclassical expressions of the two-term BCH
expansion (\ref{eq:BCH1}). 
The operator $\Sigma_t$ (\ref{eq:sigdef}) giving the first order BCH term can be computed 
as a trivial geometric series
\begin{equation}
\Sigma_t \cong \sum_{\vec{m}\neq\vec{0}}\tilde{v}_{\vec{m}}(\vec{J})
\left(1 - e^{\ii\vec{m}\cdot\vec{\omega}(\vec{J})t}\right)     
e^{\ii\vec{m}\cdot\vec{\Theta}} 
\label{eq:stsc}
\end{equation}
yielding a quasi-periodic and hence bounded temporal behaviour (due to $v_{\vec{0}}(\vec{j})=0$).
Here we have introduced modified Fourier coefficients $\tilde{v}_{\vec{m}}(\vec{j})$ 
of the perturbation 
\begin{equation}
\tilde{v}_{\vec{m}}(\vec{j}) = 
\frac{\tau v_{\vec{m}}(\vec{j})}{1-e^{\ii\vec{m}\cdot\vec{\omega}(\vec{j})}}=\ii e^{-\ii \vec{m}\cdot\vec{\omega}(\vec{j})/2}\frac{\tau v_\vec{m}(\vec{j})}
{2\sin{(\frac{1}{2}\vec{m}\cdot\vec{\omega}(\vec{j}))}}.
\label{eq:tilde}
\end{equation}
As for the operator $\bbar{V}$, or $\Gamma_t$, giving the second order BCH term, 
the calculation is more tedious. First, we plug the semiclassical dynamics 
(\ref{eq:scVt}) into the definition (\ref{eq:defbbarV}). Second, we compute the 
resulting commutators of the form
$$
[v_{\vec{m}'}(\vec{J})e^{\ii\vec{m}'\cdot\vec{\omega}(\vec{J})t'}e^{\ii\vec{m}'\cdot\vec{\Theta}},
v_{\vec{m}''}(\vec{J})e^{\ii\vec{m}''\cdot\vec{\omega}(\vec{J})t''}e^{\ii\vec{m}''\cdot\vec{\Theta}}],
$$
in the leading order in $\hbar$, by means of the Poisson brackets
$$
\ii\hbar\left(
\partial_{\vec{\theta}}(v_{\vec{m}'}e^{\ii\vec{m}'\cdot(\vec{\theta}+\vec{\omega}t')})\cdot
         \partial_{\vec{j}}(v_{\vec{m}''}e^{\ii\vec{m}''\cdot(\vec{\theta}+\vec{\omega}t'')})
         - \partial_{\vec{j}}(v_{\vec{m}'}e^{\ii\vec{m}'\cdot(\vec{\theta}+\vec{\omega}t')})\cdot
             \partial_{\theta}(v_{\vec{m}''}e^{\ii\vec{m}''\cdot(\vec{\theta}+\vec{\omega}t'')})\right)
$$
and substitution of variables $\vec{j},e^{\ii\vec{m}\cdot\vec{\theta}}$ by operators 
$\vec{J},e^{\ii\vec{m}\cdot\vec{\Theta}}$. Third, we drop the terms for which 
$\vec{m}'+\vec{m}''\neq \vec{0}$, since these contain the shift operator
$e^{\ii(\vec{m}'+\vec{m}'')\cdot\vec{\Theta}}$ [eq. (\ref{eq:shift})], 
giving off-diagonal matrix elements only which we know should give vanishing overall 
contribution for a residual observable, see eqs. (\ref{eq:gammame},\ref{eq:Vbar})\footnote[1]{Actually, 
in the off-diagonal matrix elements of the {\em Poisson bracket} we have oscillating functions of time $e^{(\vec{m}'+\vec{m}'')\cdot\vec{\omega} t}$ which, for times longer 
than $\sim 1/\hbar$, can no longer reproduce the matrix element of the {\em commutator}.}.
As a result we find the following semiclassical expression
\begin{eqnarray}
&&\bbar{V} \cong \bbar{v}(\vec{J}), \nonumber\\
&&\bbar{v}(\vec{j}) = -\frac{\tau}{2}\sum_{\vec{m}\neq 0} \vec{m}\cdot\partial_{\vec{j}}\left\{
|v_{\vec{m}}(\vec{j})|^2 \cot\left({\textstyle{\frac{1}{2}}}\vec{m}\cdot\vec{\omega}(\vec{j})\right)
\right\}.
\label{eq:bbarv}
\end{eqnarray}
\par
We have derived the semiclassical expression for both terms occurring in the echo operator 
(\ref{eq:BCH1}), for the integrated perturbation $\Sigma_t$ (\ref{eq:stsc}), and for the doubly-averaged 
perturbation $\bbar{V}$ (\ref{eq:bbarv}). 
%%% We add here a discussion on potential small denominator problem 
However, we note that both semiclassical expressions (\ref{eq:stsc},\ref{eq:bbarv})
are subject to a potential `small denominator' problem which is closely related to the one in KAM theory.
This well known problem of divergence of sums over the Fourier index $\vec{m}$ can be avoided in a generic case. 
First, strict singularities at resonances $\vec{m}\cdot \vec{\omega} = 0 \pmod{2\pi}$, where frequencies 
$\vec{\omega}$ are evaluated in the eigenstates $\ket{\vec{n}}$, happen with probability zero.
Second, the {\em near resonances} give a finite total contribution if one assumes that the classical limit of the 
perturbation $v(\vec{j},\vec{\theta})$ is sufficiently smooth, {\em e.g.} analytic in angles $\vec{\theta}$ 
such that the Fourier coefficients $v_{\vec{m}}$ fall off 
exponentially in $|\vec{m}|$. This will be assumed throughout the rest of this paper,
whereas the cases of more singular perturbations call for further investigations.
The problem is even less severe if the Fourier series (\ref{eq:fourier}) is finite as will be the
case in our numerical example.
%%% 
The results (\ref{eq:stsc},\ref{eq:bbarv}) enable us to proceed 
with the actual calculation of fidelity decay for `long' and `short' times, where the operators $\bbar{V}$ 
and $\Sigma_t$ are dominanting, respectively. 

\subsection{Asymptotic regime of long times}

Thus for sufficiently long times $t$, say longer than a certain 
$t_2 \propto \delta^{-1}$  such that the second term 
$\bbar{V} t\tau\delta^2/2$ in BCH expansion (\ref{eq:BCH},\ref{eq:BCH1}) dominates the 
first one $\Sigma_t \delta$, the fidelity (amplitude) can be written as
\begin{equation}
f(t) \cong \ave{\exp\left(\ii\frac{\tau\delta^2}{2\hbar}\bbar{V} t\right)}, \quad{\rm for}\quad t > t_2.
\label{eq:ferg}
\end{equation}
Since both operators, $\Sigma_t$ and $\bbar{V}$, have well defined
classical limits, it is clear that $t_2$ will generally {\em not depend} on
$\hbar$, however it may depend on the precise structure of the
initial state. Roughly it can be estimated semiclassically as
\begin{equation}
t_2 = \left(\frac{1}{|\bbar{v}|}\sqrt{\sum_{\vec{m}\neq\vec{0}}|\tilde{v}_{\vec{m}}|^2}\right)_{\rm ef}
\frac{2}{\tau \delta}
\label{eq:t2}
\end{equation}
where subscript ${\rm ef}$ means {\em effective value} in the action
space region of interest, {\em i.e.} where the initial state is distributed.
Note that the actual timescale $t_2$ of the dominance of the second order can be in fact
up to a factor $\sim\hbar^{-1/2}$ longer for coherent initial states and for
sufficiently small perturbation, as explained in section \ref{sec:SACS}.

The formula (\ref{eq:ferg}) can be transformed, following Ref.\cite{PZ}, 
into a very useful expression for the semiclassical analysis by, first, 
writing out the average as the trace in EBK basis $\ket{\vec{n}}$, second, 
using the fact that $\bbar{V}$ is diagonal in $\ket{\vec{n}}$ with eigenvalues 
$\cong \bbar{v}(\hbar\vec{n})$, and third, semiclassically approximating the sum $\sum_{\vec{n}}$
by an integral over the action space $\hbar^{-d}\int \d^d \vec{j}$:
\begin{eqnarray}
f(t) &\cong& \hbar^{-d} \int \d^d\vec{j} 
\exp\left(\ii\frac{\tau\delta^2}{2\hbar}\bbar{v}(\vec{j})t\right)
D_\psi(\vec{j}), \quad t_2 < t < t^*
\label{eq:ferg2}\\
D_\psi(\hbar\vec{n}) &:=& |\braket{\psi}{\vec{n}}|^2.
\label{eq:SF}
\end{eqnarray}
The last step is justified for (classically long) times up to 
$t^*$, such that the variation of the exponential in 
(\ref{eq:ferg2}) across one
Planck cell of diameter $\hbar$ is small, 
\begin{equation}
t^* = \frac{1}{|\partial_{\vec{j}}\bbar{v}|_{\rm ef}} \frac{1}{\tau \delta^2}
\sim \hbar^0 \delta^{-2}
\label{eq:tstar}
\end{equation}
We note a strong formal similarity between the action-space-integral 
(ASI) representation of fidelity for
a residual perturbing observable (\ref{eq:ferg2}) and the ASI representation for a generic observable \cite{PZ}
for times up to $\sim \hbar^0 \delta^{-1}$
\begin{equation}
f_{\rm generic}(t) \cong \hbar^{-d} \int \d^d\vec{j} 
\exp\left(\ii\frac{\tau\delta}{\hbar}\bar{v}(\vec{j})t\right)D_\psi(\vec{j}).
\label{eq:fnonerg}
\end{equation}
This means that only $\bar{v}\delta$ has to be replaced by $\bbar{v}\delta^2/2$ in the
semiclassical analysis of formula (\ref{eq:fnonerg}) elaborated in \cite{PZ}. 
This shall be discussed in detail for the specific cases of coherent and
random initial states in sections \ref{sec:SACS} and \ref{sec:SARS}.

\subsection{The plateau: linear response and beyond}
\label{sec:LRandB}

For times $t$ smaller than $t_2$ (\ref{eq:t2}) the 
first term in the exponential of (\ref{eq:BCH1}) dominates over the second
one, so we may write the fidelity amplitude generally as
\begin{equation}
f(t) \cong \ave{\exp\left(\ii\frac{\delta}{\hbar}\Sigma_t\right)}, \quad{\rm for}\quad t < t_2.
\label{eq:ferg1}
\end{equation}
Let us first discuss the regime of sufficiently small perturbation such that
the fidelity is close to $1$, {\em i.e.} the norm of the exponential is small, 
$||\delta\Sigma_t/\hbar|| \ll 1$,
so we can use the second order expansion of (\ref{eq:ferg1})
which is precisely the linear response formula (\ref{eq:LR2}).
We have to compute the uncertainty of the time integrated perturbation 
operator $\Sigma_t$.
From semiclassical expression for $\Sigma_t$ (\ref{eq:stsc}) we can directly compute the expectation value
\begin{equation}
\ave{\Sigma_t} \cong \sum_{\vec{n}}\sum_{\vec{m}\neq\vec{0}}
\tilde{v}_{\vec{m}}(\hbar\vec{n})\left(1-e^{\ii\vec{m}\cdot\vec{\omega}(\hbar\vec{n})t}\right)
\psi_{\vec{n}-\vec{m}}\psi^*_{\vec{n}},
\label{eq:EVS}
\end{equation}
where $\psi_\vec{n} := \braket{\vec{n}}{\psi}.$
Similarly, we compute the expectation value of its square
\begin{eqnarray}
\ave{\Sigma^2_t} &\cong& \sum_{\vec{n}}\sum_{\vec{m},\vec{m}'\neq \vec{0}}
\tilde{v}_{\vec{m}}(\hbar\vec{n})\tilde{v}_{\vec{m}'}(\hbar\vec{n}) \label{eq:EVS2}\\
&\times&\left(1 -
      e^{\ii\vec{m}\cdot\vec{\omega}(\hbar\vec{n})t} -
      e^{\ii\vec{m}'\cdot\vec{\omega}(\hbar\vec{n})t} +
e^{\ii(\vec{m}+\vec{m}')\cdot\vec{\omega}(\hbar\vec{n})t}
\right)
\psi_{\vec{n}-\vec{m}}\psi^*_{\vec{n}+\vec{m}'}. \nonumber
\end{eqnarray}
For sufficiently {\em many non-vanishing components} $\psi_{\vec{n}}$ and/or
for sufficiently {\em large times} $t > t_1$, 
and away from certain {\em resonance condition} (all three 
conditions will be discussed in detail later) the
terms with explicitly time-dependent oscillating factors $
\exp(\ii\vec{m}\cdot\vec{\omega}(\hbar \vec{n})t)$ can be argued to give
vanishing or (semiclassically) negligible contributions. 
In fact, the time scale $t_1$ will be determined by the condition that at typical
later times {\em random phase approximation} can be used in dealing with
the exponentials in eqs. (\ref{eq:EVS},\ref{eq:EVS2}).
Thus the above expectation values should be {\em time independent} 
and equal to their time averages
\begin{eqnarray}
\ave{\Sigma_t} &\cong& \overline{\ave{\Sigma}} = \sum_{\vec{n}}\sum_{\vec{m}\neq\vec{0}}
\tilde{v}_{\vec{m}}\psi_{\vec{n}-\vec{m}}\psi^*_{\vec{n}}, \label{eq:EVSA}\\
\ave{\Sigma_t^2} &\cong& \overline{\ave{\Sigma^2}} = 
\sum_{\vec{n}}\left\{\Bigl|\sum_{\vec{m}\neq 0}\tilde{v}_{\vec{m}}(\hbar\vec{n})\psi_{\vec{n}-\vec{m}}\Bigr|^2
+ \sum_{\vec{m}\neq 0}\left|\tilde{v}_{\vec{m}}(\hbar\vec{n})\psi_{\vec{n}-\vec{m}}\right|^2\right\}.
\label{eq:EVSA2}
\end{eqnarray}
This gives us a prediction that, after following a classical decay up to 
the short time $t_1$, the fidelity should reach
a constant value -- a plateau and stay there up to time $t_2$. The linear response value of the fidelity at the plateau is
\begin{equation}
F(t) \cong 1 - \frac{\delta^2}{\hbar^2}\left\{
\overline{\ave{\Sigma^2}} - \overline{\ave{\Sigma}}^2\right\} + 
{\cal O}(\delta^4),\quad
{\rm for}\quad t_1 < t < t_2.
\label{eq:plateauLR}
\end{equation}

Further, we can easily go beyond the linear response approximation
by expanding the formula (\ref{eq:ferg1}) to all orders in $\delta$.
For this, we have to calculate the powers of the operator $\Sigma_t$
\begin{equation}
\Sigma_t^k \cong \sum_{\vec{m}_1,\ldots,\vec{m}_k\neq\vec{0}}
\prod_{l=1}^k \tilde{v}_{\vec{m}_l}(\vec{J})e^{\ii\vec{m}_l\cdot\vec{\Theta}}
\left(1-e^{\ii\vec{m}_l\cdot\vec{\omega}(\vec{J})t}\right).
\label{eq:sigprod}
\end{equation}
We shall use two facts in order to carry through the calculation: 
(i) That for leading semiclassical order the operator ordering is not important. 
(ii) Oscillatory time dependent terms, which in any case average to zero, typically 
give exponentially damped or semiclassically small overall contributions 
when used in expectation values. Thus we shall approximate $\Sigma_t^k$ by its time-average $\overline{\Sigma^k}$, and the latter is calculated by selecting
from the product (\ref{eq:sigprod}) only the combinations of multi-indices which
sum up to zero $\vec{m}_1 + \cdots + \vec{m}_l = \vec{0}$.
Let us write the time average of the operator $\Sigma_t$ in terms of an
explicit function of canonical operators
\begin{eqnarray}
&& \overline{\Sigma} \cong \tilde{v}(\vec{J},\vec{\Theta}), \nonumber\\ 
&& \tilde{v}(\vec{j},\vec{\theta}) := 
\sum_{\vec{m}\neq\vec{0}}\tilde{v}_{\vec{m}}(\vec{j})e^{\ii\vec{m}\cdot\vec{\theta}}.
\label{eq:vtilde}
\end{eqnarray}
Then we calculate
\begin{eqnarray}
\overline{\Sigma^k} &\cong&
\overline{\Sigma}^k + \sum_{l=2}^k (-1)^l {\textstyle{k\choose l}}
\overline{\Sigma}^{k-l}
\!\!\!\!\!\!\sum_{\vec{m}_1,\ldots,\vec{m}_l\neq \vec{0}}\!\!\!\!
\delta_{\vec{m}_1+\cdots+\vec{m}_l}
\tilde{v}_{\vec{m}_1}(\vec{J})\cdots
\tilde{v}_{\vec{m}_l}(\vec{J}) \nonumber\\
&=& \frac{1}{(2\pi)^d} 
\sum_{l=0}^k (-1)^l {\textstyle{k\choose l}}
\tilde{v}(\vec{J},\vec{\Theta})^{k-l} 
\int\d^d \vec{x} [\tilde{v}(\vec{J},\vec{x})]^l.
\label{eq:binomial}
\end{eqnarray}
Please observe that $\vec{\Theta}$ is an angle-operator (which always stands in the
exponential, so it is well defined), and $\vec{x}$ is a $d-$dimensional
integration variable, and that in order to write (\ref{eq:binomial})
we have used an integral representation of the Kronecker symbol: 
$\delta_{\vec{m}} = (2\pi)^{-d}\int\d^d\vec{x}
e^{\ii\vec{m}\cdot\vec{x}}$.
Now it is straightforward to compute the power series 
$\sum_{k=0}^\infty(\ii \delta/\hbar)^k \overline{\Sigma^k}/k!$ by changing the 
summation variables to $k-l$ and $k$, yielding a product of two exponentials
\begin{equation}
\overline{\exp\left(\frac{\ii\delta}{\hbar}\Sigma\right)}\cong
\exp\left(\frac{\ii\delta}{\hbar}\tilde{v}(\vec{J},\vec{\Theta})\right)
\int \frac{\d^d\vec{x}}{(2\pi)^d} \exp\left(-\frac{\ii\delta}{\hbar}\tilde{v}(\vec{J},\vec{x})\right),
\label{eq:expaverage}
\end{equation}
and the plateau of fidelity (amplitude) is the expectation value of this operator
\begin{equation}
f_{\rm plateau} \cong 
\int\frac{d^d\vec{x}}{(2\pi)^d}
\ave{\exp\left(\frac{\ii\delta}{\hbar}\tilde{v}(\vec{J},\vec{\Theta})\right)
\exp\left(\!-\frac{\ii\delta}{\hbar}\tilde{v}(\vec{J},\vec{x})\right)},
\label{eq:plateau}
\end{equation}
$f(t) = f_{\rm plateau}$ for $t_1 < t < t_2$.
The very existence of such a plateau of fidelity (high fidelity for small
$\delta/\hbar$) is very interesting and 
distinct property of quantum dynamics. Note that the time scale $t_1$ only
depends on the unperturbed dynamics, namely on the property of the
operator $\Sigma_t$, so it cannot depend on the strength of the perturbation, 
$t_1 = {\cal O}(\delta^0)$. Thus the range of the plateau, {\em i.e.}
$t_2/t_1 \propto \delta^{-1}$ can become arbitrarily large for small $\delta$. 
The formula (\ref{eq:plateau}) becomes very useful whenever one is able to semiclassically
compute the quantum expectation value in terms of classical phase space integrals.
We shall present this derivation for two extremal cases of {\em coherent} and {\em random} 
initial states, in sections \ref{sec:SACS} and \ref{sec:SARS}, respectively.
%In these cases the phase space integrals are elementary and can be expressed 
%in terms of Bessel functions.
\par
Note that the formulae (\ref{eq:expaverage},\ref{eq:plateau}) may be very useful in a general 
case whenever one has to calculate an expectation value of the form 
$\ave{\exp\left(\ii\frac{\delta}{\hbar}\Sigma_t\right)}$ where 
$\Sigma_t$ is a time-integrated quasi-periodic process with a zero time-average.

\section{Numerical example: integrable top}

For numerical illustration of the above theory we take a spin system with the
following one-time-step unitary propagator
\begin{equation}
U_0=\exp{\left\{-\ii S\frac{\alpha}{2} \left(\frac{S_{\rm z}}{S} -\beta\right)^2\right\}},
\label{eq:Ktop}
\end{equation}
with parameters $\alpha$ and $\beta$. $S_k, k={\rm x, y, z}$ are standard quantum angular momentum operators with a fixed magnitude $S$ of angular momentum 
and with the SU(2) commutator $[S_k,S_l]=\ii\varepsilon_{klm}S_m$. 

The semiclassical limit is obtained by letting $S\to\infty$ while the classical angular momentum $\hbar S=1$ is kept fixed, so that the effective Planck constant is given by $\hbar=1/S$. 
The classical map corresponding to the one-time-step propagator $U_0$ can be obtained from the 
Heisenberg equations of angular momentum operators in the $S\to\infty$ limit. Defining by 
$(x,y,z)=(S_{\rm x},S_{\rm y},S_{\rm z})/S$ a point on a unit sphere, 
we obtain a classical area preserving map:
\begin{eqnarray}
x_{t+1}&=&x_t \cos{(\alpha (z_t-\beta))}-y_t\sin{(\alpha (z_t-\beta))} \nonumber \\
y_{t+1}&=&y_t \cos{(\alpha (z_t-\beta))}+x_t\sin{(\alpha (z_t-\beta))} \\
z_{t+1}&=&z_t. \nonumber
\label{eq:Cmap}
\end{eqnarray}
This classical map represents a twist around ${\rm z}$-axis. We note that it corresponds to the
stroboscopic map (\ref{eq:hamileq}) with an arbitrary unit of time, so we put $\tau=1$, of an integrable
system with the Hamiltonian $h_0(j) = \frac{1}{2}\alpha (j-\beta)^2$ generating a frequency field
\begin{equation}
\omega(j) = \frac{\d h_0(j)}{\d j} = \alpha (j-\beta).
\end{equation}
Here we used a canonical transformation from a unit-sphere to an action-angle pair 
$(j,\theta)\in [-1,1]\times [0,2\pi)$, namely
\begin{equation}
x = \sqrt{1-j^2}\cos\theta,
\qquad
y = \sqrt{1-j^2}\sin\theta,
\qquad
z = j.
\qquad
\label{eq:can}
\end{equation}
Now we perturb the Hamiltonian by periodic kicking with a transverse pulsed magnetic field in $x$ 
direction,
\begin{equation}
h_\delta(j,\theta,\tau) = \frac{1}{2}\alpha (j-\beta)^2 + 
\delta \sqrt{1-j^2}\cos\theta \sum_{k=-\infty}^\infty \delta(\tau-k).
\end{equation}
Perturbed quantum evolution is given by a product of two unitary propagators
\begin{equation}
U_\delta = U_0\exp{(-\ii \delta S_{\rm x})},
\label{eq:Jx}
\end{equation}  
so the perturbation generator is 
\begin{equation}
V = S_{\rm x}/S.
\label{eq:V}
\end{equation}
The classical perturbation has only one Fourier component, namely
\begin{equation}
v(j,\theta) = \sqrt{1-j^2}\cos\theta,\quad v_{\pm 1}(j) = \frac{1}{2}\sqrt{1-j^2},
\label{eq:VFour}
\end{equation}
whereas $v_0\equiv 0$ indicating that the time-average vanishes $\bar{v}=0$, and 
$\bar{V}=0$. 
\par
In our numerical illustrations two different types of initial states will be used. 
SU(2) coherent wavepackets are used to probe the correspondence with the classical fidelity, 
while random states are used to investigate the other end -- states without a classical 
correspondence. The parameter $\alpha$ in $U_0$ (\ref{eq:Ktop}) will be always set to
$\alpha=1.1$, while $\beta=0$ for coherent initial states, and 
$\beta=0$ and $\beta=1.4$ for random initial states. The reason for choosing nonzero shift
$\beta$ for random states will be explained later. We should stress that we have performed calculations also 
for other choices of regular $U_0$, also in KAM regime, {\em e.g.} for precisely the same model and 
parameter values as used in Ref.~\cite{wrongpaper},
and obtained qualitatively the same results as for the presented case of unperturbed dynamics.  
The coherent state written in the canonical eigenbasis $\ket{m}$ of the operator $S_{\rm z}$ and centered at the position 
$\vec{n}=(\sin{\vartheta^*}\cos{\varphi^*},\sin{\vartheta^*}\sin{\varphi^*},\cos{\vartheta^*})$ 
on a unit sphere is
\begin{equation}
\ket{\vartheta^*,\varphi^*}=\sum_{m=-S}^{S}{\left( {2S \atop S+m} \right)^{1/2} 
\cos{(\vartheta^*/2)}^{S+m} \sin{(\vartheta^*/2)}^{S-m} e^{-\ii m \varphi^*} \ket{m}}.
\label{eq:SU2coh}
\end{equation}
The corresponding classical density reads \cite{Fox}
\begin{equation}
\rho_{\rm cl}(\vartheta,\varphi)=\frac{4S+1}{4\pi}\exp{\left\{-S\left((\vartheta-\vartheta^*)^2+(\varphi-\varphi^*)^2\sin^2{\vartheta}\right)\right\} }.
\label{eq:ClasSU2}
\end{equation}
In the numerical experiments reported below the coherent initial state will always 
be positioned at the point $(\vartheta^*,\varphi^*)=(1,1)$.

\section{Semiclassical asymptotics: coherent initial state}
\label{sec:SACS}

Let us now study an important specific case of a (generally squeezed) 
coherent initial state which can be written in the EBK basis as a general 
Gaussian centered around a phase space point $(\vec{j}^*,\vec{\theta}^*)$
\begin{equation}
\!\!\!\!\!\!\!\!\!
\braket{\vec{n}}{{\vec{j}^*,\vec{\theta}^*}} \cong
\left(\frac{\hbar}{\pi}\right)^{d/4}
\!\!\!\left|\det\Lambda\right|^{1/4}
\exp\left\{-\frac{1}{2\hbar}(\hbar\vec{n}\!-\!\vec{j}^*)\!\cdot\!\Lambda(\hbar\vec{n}\!-\!\vec{j}^*) - 
\ii\vec{n}\!\cdot\!\vec{\theta}^*\right\},
\label{eq:CS}
\end{equation}
with $\Lambda$ being a positive symmetric $d\times d$ matrix of squeezing 
parameters. Note that the shape of the coherent state is generally only 
asymptotically Gaussian, as $\hbar\to 0$, due to cyclic and discrete 
nature of coordinates $\vec{\theta}$ and $\vec{j}$, respectively.
Let us also write the {\em structure function} (\ref{eq:SF}) of our coherent 
state (\ref{eq:CS})
\begin{equation}
D_\psi(\vec{j}) \cong \left(\frac{\hbar}{\pi}\right)^{d/2}
\!\!\!\left|\det\Lambda\right|^{1/2}
\exp\left\{-\frac{1}{\hbar}(\vec{j}\!-\!\vec{j}^*)\!\cdot\!\Lambda(\vec{j}\!-\!\vec{j}^*)\right\}
\label{eq:CSSF}
\end{equation}
which is normalized as $\hbar^{-d}\int \d^d\vec{j} D_\psi(\vec{j}) = 1$.

For example, for SU(2) coherent state of a quantum top (\ref{eq:SU2coh}) when written in the asymptotic form (\ref{eq:CS}), the squeezing parameter reads $\Lambda = 1/\sin^2 \vartheta^*$.

\subsection{The plateau: linear response and beyond}

\begin{figure}[ht]
\centerline{\includegraphics{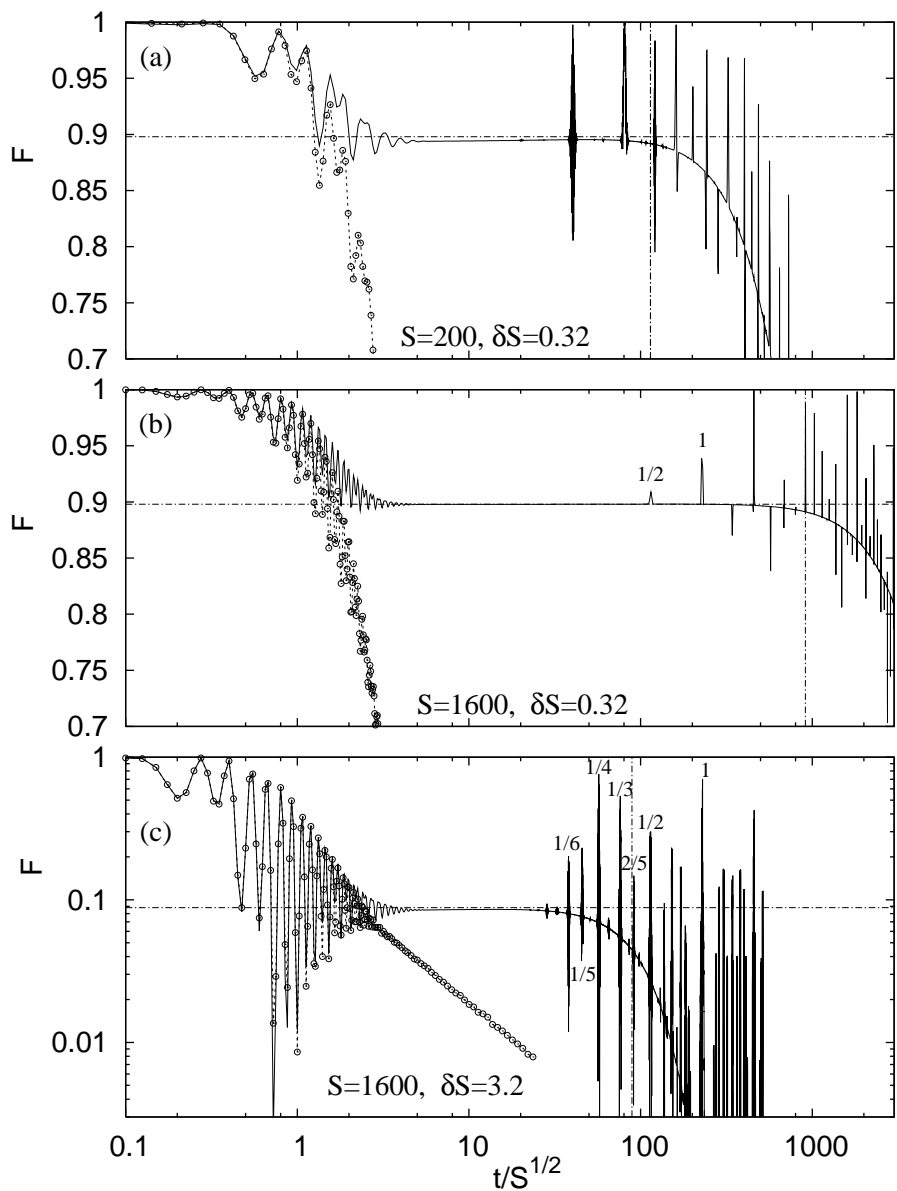}}
\caption{The short time fidelity for a quantized top $\alpha=1.1,\beta=0$ is shown for the coherent initial state,
for $S=200$ (a), and $S=1600$ (b), with a fixed product $\delta S=0.32$
being well described by the linear response. In (c) we show $S=1600$ and stronger perturbation with 
$\delta S = 3.2$.
Note that the time axis is rescaled as $t/t_1$.
Symbols connected with dashed lines denote the corresponding classical fidelity. The horizontal chain line 
denotes the theoretical value of the plateau (\ref{eq:Plat}), while the vertical chain line denotes the estimated 
theoretical value for $t_2$ (\ref{eq:t2coh}). In (b,c) we also indicate fractional $2\pi k/p$ resonances with 
$k/p$ marked on the figure (see text for details).} 
\label{fig:shortCoh}
\end{figure}   

In the regime of linear response, valid for sufficiently small $\delta$, we
simply evaluate the general expressions (\ref{eq:EVS},\ref{eq:EVS2}) for the
particular case of a coherent initial state (\ref{eq:CS}), namely we write
$\psi_{\vec{n}} = \braket{\vec{n}}{\vec{j}^*,\vec{\theta}^*}$.
First we will show that the time dependent terms in expectation values of the powers of $\Sigma_t$ indeed vanish for $t>t_1$ as stated in section \ref{sec:LRandB}. 
We recall the assumption that the perturbation $v(\vec{j},\vec{\theta})$ is
sufficiently smooth, {\em e.g.} analytic in $\vec{\theta}$, so that the
Fourier coefficients $v_{\vec{m}}(\vec{j})$ decrease sufficiently fast,
{\em e.g.} exponentially, or only a finite number of $v_{\vec{m}}(\vec{j})$ 
is non-vanishing. What we actually need here is that an effective number of Fourier components
is smaller than the width of a wavepacket which is $\sim \hbar^{-1/2}$.
This means that within the range of Fourier series over 
$\vec{m}$, or $\vec{m}'$ we can approximate
\begin{equation}
\psi_{\vec{n}-\vec{m}}^* \psi_{\vec{n}+\vec{m}'} \cong
D_\psi(\hbar\vec{n}) e^{-\ii(\vec{m}+\vec{m'})\cdot\vec{\theta}^*}.
\end{equation}
Let us estimate the general time dependent term of expressions 
(\ref{eq:EVS},\ref{eq:EVS2}), where all factors with non-singular classical limit are combined 
together and denoted as $g(\vec{j})$, by means of expanding the frequency around the center of the packet
$\vec{\omega}(\vec{j}^* + \vec{x}) = \vec{\omega}(\vec{j}^*) + \Omega \vec{x} + \ldots$, where $\Omega$ is a matrix
$\Omega_{kl} = \partial\omega_k(\vec{j}^*)/\partial j_l$, followed by $d$-dimensional Gaussian integration
\begin{eqnarray}
&&\sum_{\vec{n}} g(\hbar\vec{n})e^{\ii\vec{m}\cdot\vec{\omega}(\hbar\vec{n})t} D_\psi(\hbar\vec{n})
\cong\hbar^{-d}\int\d^d\vec{j} g(\vec{j}) e^{\ii\vec{m}\cdot\vec{\omega}(\vec{j})t} D_\psi(\vec{j})\nonumber\\
&&\cong g(\vec{j}^*)e^{\ii\vec{m}\cdot\vec{\omega}(\vec{j}^*)t} \left(\frac{\hbar}{\pi}\right)^{d/2}
\!\!\!\left|\det\Lambda\right|^{1/2}
\int\d^d \vec{x} \exp\left(-\frac{1}{\hbar} \vec{x}\cdot \Lambda \vec{x} + \ii t \vec{m}\cdot\Omega\vec{x}\right)
\nonumber\\
&&= g(\vec{j}^*)e^{\ii\vec{m}\cdot\vec{\omega}(\vec{j}^*)t}
\exp\left(-\frac{\hbar t^2}{4}\vec{m}\cdot \Omega \Lambda^{-1} \Omega^T\vec{m}\right).
\label{eq:gauss}
\end{eqnarray}
We see that all these terms decay to zero with Gaussian envelopes with the longest time scale
estimated as
\begin{equation}
t_1 = \left( \frac{\hbar}{4} \min_{\vec{m}\neq\vec{0}}\left(\vec{m}\cdot \Omega \Lambda^{-1}\Omega^T\vec{m}\right)\right)^{-1/2}
\propto \hbar^{-1/2}.
\label{eq:t1}
\end{equation}
Note that the decay of (\ref{eq:gauss}) is absent if $\Omega=0$, {\em e.g.} in the case 
of $d$-dimensional harmonic oscillator. There may also be a general problem 
with the formal existence of the scale $t_1$
(\ref{eq:t1}) if the derivative matrix $\Omega$ is singular, but this may not actually affect the fidelity for 
sufficiently fast converging or finite Fourier series (\ref{eq:fourier}). 

Thus we have shown that for $t > t_1$ expectation values 
(\ref{eq:EVS},\ref{eq:EVS2}) are indeed given by time-independent 
expressions (\ref{eq:EVSA},\ref{eq:EVSA2}) which in the case of a 
coherent initial state (\ref{eq:CS}) evaluate to
\begin{eqnarray}
\ave{\Sigma_t} &\cong& \sum_{\vec{m}\neq\vec{0}} \tilde{v}_{\vec{m}}(\vec{j}^*)e^{\ii\vec{m}\cdot\vec{\theta}^*} = \tilde{v}(\vec{j}^*,\vec{\theta}^*), 
\label{eq:EVSCS}\\
\ave{\Sigma^2_t} &\cong&  
\left(\tilde{v}(\vec{j}^*,\vec{\theta}^*)\right)^2
+ \sum_{\vec{m}\neq 0}\left|\tilde{v}_{\vec{m}}(\vec{j}^*)\right|^2.
\label{eq:EVSCS2}
\end{eqnarray}
The variance $\nu=\ave{\Sigma_t^2}-\ave{\Sigma_t}^2$ which determines the plateau in the fidelity (\ref{eq:LR2})
is the second term on RHS of eq. (\ref{eq:EVSCS2}). In terms of original Fourier coefficients (\ref{eq:tilde})
the final linear response result reads
\begin{eqnarray}
&&F_{\rm coh}(t) = 1 - \frac{\delta^2}{\hbar^2}\nu_{\rm coh} + {\cal O}(\delta^4),\quad t > t_1,\nonumber\\
&&\nu_{\rm coh}=\sum_{\vec{m}\neq\vec{0}}\frac{\tau^2 |v_{\vec{m}}(\vec{j}^*)|^2}{4\sin^2(\frac{1}{2}\vec{m}\cdot\vec{\omega}(\vec{j}^*))}.
\label{eq:LRCS}
\end{eqnarray}

Beyond the linear response approximation, the value of the plateau can be
computed by applying a general formula for $f_{\rm plateau}$ (\ref{eq:plateau}). We shall make use 
of the fact that for coherent states we have the expectation value
\begin{equation}
\ave{\exp(-(\ii\delta/\hbar)g(\vec{J},\vec{\Theta}))} \cong
\exp(-(\ii\delta/\hbar)g(\vec{j}^*,\vec{\theta}^*)),
\end{equation}
for some smooth function $g$, provided that the diameter of the wavepacket
$\sim\sqrt{\hbar}$ is smaller than the oscillation scale of the
exponential $\sim \hbar/\delta$, {\em i.e.} provided $\delta \ll \hbar^{1/2}$.
Then the squared modulus of $f_{\rm plateau}$ (\ref{eq:plateau}) rewrites as
\begin{equation}
F_{\rm plateau} \cong \frac{1}{(2\pi)^{2d}}
\left|\int \d^d\vec{x} \exp\left(-\frac{\ii\delta}{\hbar}\tilde{v}(\vec{j}^*,\vec{x})\right)\right|^2.
\label{eq:plateauCS}
\end{equation}
The expression for $\nu_{\rm coh}$ (\ref{eq:LRCS}) is of course just the lowest order expansion of 
$F_{\rm plateau}$ (\ref{eq:plateauCS}). It is interesting to note that the angle $\vec{\theta}^*$ 
does not affect the probability $F_{\rm plateau}$ as it only rotates the phase of the amplitude 
$f_{\rm plateau}$.

For smaller times $t < t_1 \propto \hbar^{-1/2}$ the quantum fidelity is expected to 
follow the classical fidelity as defined by the overlap of two initially 
Gaussian classical phase space 
densities evolved under slightly different quasi-regular time evolutions 
[see Ref. \cite{PZ} for a definition and linear response
treatment of the classical fidelity]. 
More precisely, the quantum fidelity can be written in two equivalent ways as
\begin{eqnarray}
F(t) &=& (2\pi\hbar)^d \int\d^d\!\vec{q}\,\d^d\!\vec{p}\,W_{U_0^t\ket{\psi}}(\vec{q},\vec{p})W_{U_\delta^t\ket{\psi}}(\vec{q},\vec{p})\\ &=& 
(2\pi\hbar)^d \int\d^d\!\vec{q}\,\d^d\!\vec{p}\,W_{\ket{\psi}}(\vec{q},\vec{p}) 
W_{M_\delta(t)\ket{\psi}}(\vec{q},\vec{p}) \label{eq:Wigecho}
\end{eqnarray} 
where $W_{\ket{\psi}}(\vec{q},\vec{p})$ is 
the {\em Wigner function} of some state $\ket{\psi}$.
The corresponding classical fidelity is defined by the same formula if 
$(2\pi\hbar)^{d/2} W_{U^t\ket{\psi}}(\vec{q},\vec{p})$
is substituted by the evolving classical phase space density, a solution of 
the corresponding classical Liouville equation, with the 
initial condition $(2\pi\hbar)^{d/2} W_{\ket{\psi}}(\vec{q},\vec{p})$ 
which is proportional to the Wigner function 
of the initial state 
$\ket{\psi}$.
Of course, this only makes sense if the function
$W_{\ket{\psi}}(\vec{q},\vec{p})$ is
strictly nonnegative so that it corresponds to some classical state,
such as for example for a coherent state where it is a Gaussian.
Indeed, time $t_1 \sim \hbar^{-1/2}$ may also be interpreted as the
{\em integrable Ehrenfest time} up to which phase space point-like quantum-classical correspondence 
will hold. Namely it is consistent with the time needed for a minimal 
uncertainty wavepacket of diameter $\sim \hbar^{1/2}$ to spread 
ballistically over a region of 
the classical size ($\sim\hbar^0$) of an invariant torus. 
After this time, quantum wavepacket
will start to coherently interfere with itself, {\em e.g.} its Wigner function will develop negative values, 
so the strict quantum-classical correspondence will stop (see subsection \ref{subsect:wig}). 
Therefore we expect initial agreement between the classical and the quantum fidelity up to time $t_1$ and after that
the classical fidelity of a regular dynamics with a residual perturbation decays with a power law $\propto (\delta \hbar^{-1/2} t)^{-d}$ (factor $\hbar^{1/2}$ comes from the size of the corresponding classical
density)
[see Refs.\cite{Veble,Bruno}] whereas the quantum fidelity {\em freezes to a constant value} as computed by our semiclassical theory (\ref{eq:plateauCS}).

\begin{figure}[h!]
\centerline{\includegraphics{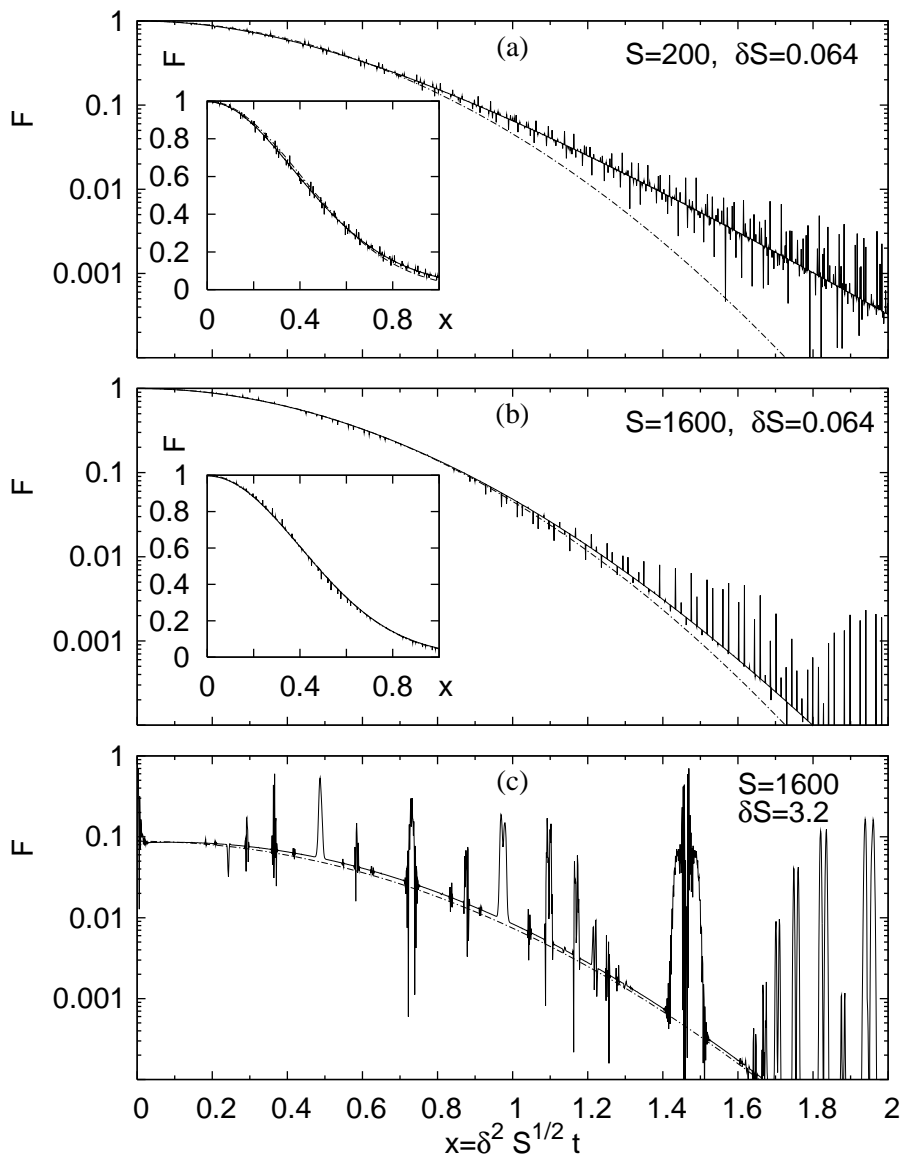}}
\caption{Long time ballistic decay of the fidelity for a quantized top with $\alpha=1.1,\beta=0$ 
and coherent initial state is shown 
for cases $S=200$ (a), and $S=1600$ (b), of weak perturbation $\delta S=0.064$,
and for strong perturbation $S=1600, \delta S=3.2$ (c).
Chain curves indicate theoretical Gaussian (\ref{eq:ADCS}) with analytically computed coefficients,
except in the case (c), where we multiply the theoretical Gaussian decay by a prefactor $0.088$ being equal to 
the theoretical value of the plateau (\ref{eq:Plat}), and rescale the exponent of the Gaussian by a factor $0.8$
taking into account the effect of non-small first term in the exponent of (\ref{eq:BCH1}). 
Note that in the limit $S \to \infty$ the agreement with the semiclassical theory improves and that the size of 
the resonant spikes is of the same order as the drop in the linear response plateau. The insets show the
data and the theory in the normal scale.}
\label{fig:longCoh}
\end{figure}
This picture is nicely confirmed by the numerical experiment with a quantized integrable top (\ref{eq:Ktop})
as shown in figure \ref{fig:shortCoh}, where we choose zero shift $\beta=0$. 
Agreement between the classical and quantum fidelity up to 
$t_1 \sim \sqrt{S}$ ($\hbar=1/S$) can be nicely observed. After $t_1$ the fidelity stays constant until $t_2$, the point 
where fidelity again starts to decrease. This second timescale $t_2$ will be discussed in the next subsection. The value of the plateau can be calculated specifically for our model by means of the semiclassical expression for $F_{\rm plateau}$ (\ref{eq:plateauCS}) and using Fourier modes of our numerical model (\ref{eq:VFour}). We get $\tilde{v}(j,\theta) = -\frac{1}{2}\sqrt{1-j^2}\sin(\theta-\frac{1}{2}\alpha j)/\sin{(\alpha j/2)}$ and the integral occurring in $F_{\rm plateau}$ is elementary and gives
\begin{equation}
F_{\rm plateau}=J_0^2\left(\delta S \frac{\sqrt{1-j^{*2}}}{2\sin{(\alpha j^*/2)}}\right),
\label{eq:Plat}
\end{equation}
with $J_0$ being the zero order Bessel function. 
Agreement with this theory is excellent both in the linear response 
regime (figure \ref{fig:shortCoh}a,b) and also for strong perturbation (figure \ref{fig:shortCoh}c). 
Observe also a power law decay of the classical fidelity $F_{\rm cl} \sim (t \delta/\hbar)^{-1}$ beyond the
regular Ehrenfest time $t > t_1$ for strong perturbation in figure \ref{fig:shortCoh}c. 
\par
Actually the calculation of $F_{\rm plateau}$ can be generalized to any 
perturbation with a single nonzero Fourier mode $\pm \vec{m}_0$ with the result
\begin{equation}
F_{\rm plateau}=J_0^2\left(\frac{\tau\delta}{\hbar}\frac{|v_{\vec{m}_0}(\vec{j}^*)|}{\sin{\left\{\vec{m}_0\cdot\vec{\omega}(\vec{j}^*)/2\right\}}}\right),
\end{equation}
whereas for a more general multi-mode perturbations we have to evaluate the integral 
(\ref{eq:plateauCS}) numerically.

\subsection{Asymptotic regime of long times}

After a sufficiently long time $t_2$ of order $\delta^{-1}$ the second order term in BCH expansion 
(\ref{eq:BCH}) will start to dominate over the
echo operator so the fidelity can be computed with our semiclassical formula (\ref{eq:ferg2}).
The straightforward calculation follows exactly the one for a generic perturbation in 
Ref.\cite{PZ}, paragraph 2.2.2, where $\bar{v}\delta$ has to be replaced by 
$\bbar{v}\delta^2/2$ so we shall not repeat it here. Such
a Gaussian approximation is justified provided the stationary point of the exponent is not moved
appreciably from the center $\vec{j}^*$ of the packet (\ref{eq:CSSF}).
This implies that $t\delta^2 \ll 1$ which is the same condition as
required by substituting the sum over quantum numbers by the integral over
the action space (\ref{eq:ferg2}). The final result reads
\begin{equation}
f_{\rm coh}(t) \cong \exp\left(-(\vec{u}\cdot\Lambda^{-1}\vec{u})\frac{\delta^4 t^2}{16\hbar} +
\frac{\ii\bbar{v}(\vec{j}^*)\delta^2\tau t}{2\hbar}\right),\qquad
\vec{u} := \tau\frac{\partial\bbar{v}(\vec{j}^*)}{\partial\vec{j}},
\label{eq:uu}
\end{equation}
where the vector $\vec{u}$ is just a gradient of the classical observable $\bbar{v}$ at the center 
of the wavepacket. Thus we have derived a Gaussian decay of the fidelity 
\begin{equation}
F_{\rm coh}(t) = \exp\left\{-\left(\frac{t}{t_{\rm coh}}\right)^2\right\},\qquad
t_{\rm coh} = (\vec{u}\cdot\Lambda^{-1}\vec{u})^{-1/2}\frac{(8\hbar)^{1/2}}{\delta^2}
\label{eq:ADCS}
\end{equation}
on a timescale $t_{\rm coh} \propto \hbar^{1/2}\delta^{-2}$.
Indeed, it can now be checked that in the semiclassical regime of small
$\hbar$ the fidelity decays well before the time limit (\ref{eq:tstar}), $t^*\sim\hbar^0\delta^{-2}$,
of our approximations.
The formula (\ref{eq:ADCS}) can only be expected to be accurate provided the plateau is
close to $1$ and hence described within the linear response approximation. Only in such a
case can the effect of the first term $\Sigma_t$ in the exponential of the echo operator (\ref{eq:BCH1}) really be neglected,
in the opposite case we can correct the Gaussian (\ref{eq:ADCS}) by multiplying it with the 
plateau value and adjusting the coefficient in the exponential (see {\em e.g.} 
figure \ref{fig:longCoh}c).

In such a {\em regime of small perturbation}, $\delta < \nu_{\rm coh}^{-1/2} \hbar$, 
we determine the crossover time $t_2$ by comparing the linear response formula (\ref{eq:LRCS}) 
with the decay law (\ref{eq:ADCS}), namely 
$1 - (t_2/t_{\rm coh})^2 = 1 - \delta^2 \nu_{\rm coh}/\hbar^2$. 
For stronger perturbation, namely up to $\delta\sim \sqrt{\hbar}$, time scale $t_2$ can be simply 
estimated by $t_{\rm coh}$, so we have a uniform estimation 
\begin{equation}
t_2 = \min\{1,\frac{\delta}{\hbar}\nu_{\rm coh}^{1/2}\} t_{\rm coh} = 
\min\{{\rm const}\ \hbar^{-1/2}\delta^{-1},{\rm const}\ \hbar^{1/2}\delta^{-2}\}
\label{eq:t2coh}
\end{equation}
We note that the crossover time $t_2$, for coherent initial states and for a small
perturbation $\delta < \nu_{\rm coh}^{-1/2} \hbar$, is in fact by a factor
$\hbar^{-1/2}$ longer than the estimate (\ref{eq:t2}). 
This is due to the fact that coherent states
are strongly localized in action coordinates (quantum numbers) for small $\hbar$. Therefore, for small 
$\delta$, the operator $\bbar{V}t\delta^2$, although it may already be dominating 
$\Sigma_t \delta$ in norm, will only 
effectively rotate the overall phase of a coherent initial state since it is diagonal in 
$\ket{\vec{n}}$ and thus will not (yet) affect the fidelity.
So the estimate (\ref{eq:t2}) is expected to be valid only for initial states whose relative
support in quantum number lattice is not shrinking as $\hbar\to 0$.
It is interesting that for the strongest allowed perturbation 
$\delta_{\rm max} = {\rm const}\ \hbar^{1/2}$ for our semiclassical theory to be valid, the 
estimate (\ref{eq:t2coh}) agrees with the general estimate (\ref{eq:t2}), $t_2 \sim 1/\delta$.

Timescale $t_2$ can be seen in figure \ref{fig:shortCoh} as the point of departure of 
fidelity from the plateau value. Using our model and the position of initial coherent state 
this can be calculated to be $t_2= \min\{0.57 \sqrt{S}/\delta, 0.57/(\delta^2 \sqrt{S}) \}$ (similarity of numerical prefactors is just a coincidence) 
which is shown with 
vertical chain lines in figure \ref{fig:shortCoh}. The theoretical position of $t_2$ is shown with a vertical chain line and is given by $t_{\rm coh}$ for a strong perturbation $\delta S=3.2$ in 
figure \ref{fig:shortCoh}c while it is $t_{\rm coh}\nu_{\rm coh}^{1/2}\delta/\hbar$ in figures \ref{fig:shortCoh}a,b. The long time decay of fidelity is shown in 
figure \ref{fig:longCoh}. Theoretical Gaussian decay (\ref{eq:ADCS}), shown with a chain curve, 
is again confirmed with an analytical formula for the decay time 
$t_{\rm coh}=0.57 \delta^{-2} \hbar^{1/2}$ evaluated at the particular position of the packet.
Note that we do not have any fitting parameters, except in the
case of a strong perturbation ($\delta S \gg 1$, figure \ref{fig:longCoh}c) where the
prefactor and the exponent of a Gaussian had to be slightly adjusted due to
the non-negligible effect of the first term in (\ref{eq:BCH1}) [see caption for details].
Quite prominent feature in figures \ref{fig:shortCoh} and \ref{fig:longCoh} are 
also ``spikes'' occurring at regular intervals, where the fidelity 
suddenly increases or wildly oscillates. These will be called the {\em echo resonances} and are 
particular to one-dimensional systems.
\par
We should remark that, although we obtain asymptotically Gaussian decay of fidelity for a 
single coherent initial state, one may be interested in an {\em effective fidelity} 
averaged with respect to phase space positions of initial coherent state \cite{wrongpaper}.
In such a case one may typically get a power law decay due to possible points in the phase space 
where the theoretical expression for $t_{\rm coh}$ diverges 
(at the positions of zeroes of $\vec{u}(\vec{j}^*)$ of eq. 
(\ref{eq:uu})), but still on a time scale $\propto\delta^{-2}$. Note that this 
{\em effective power law decay} is a general scenario and is not particular to the case of 
vanishing time-average perturbation (in the case of $\bar{V}\neq 0$ the decay time scales as 
$\delta^{-1}$).  

\subsection{Echo resonances in one dimension}

\begin{figure}[h!]
\centerline{\includegraphics{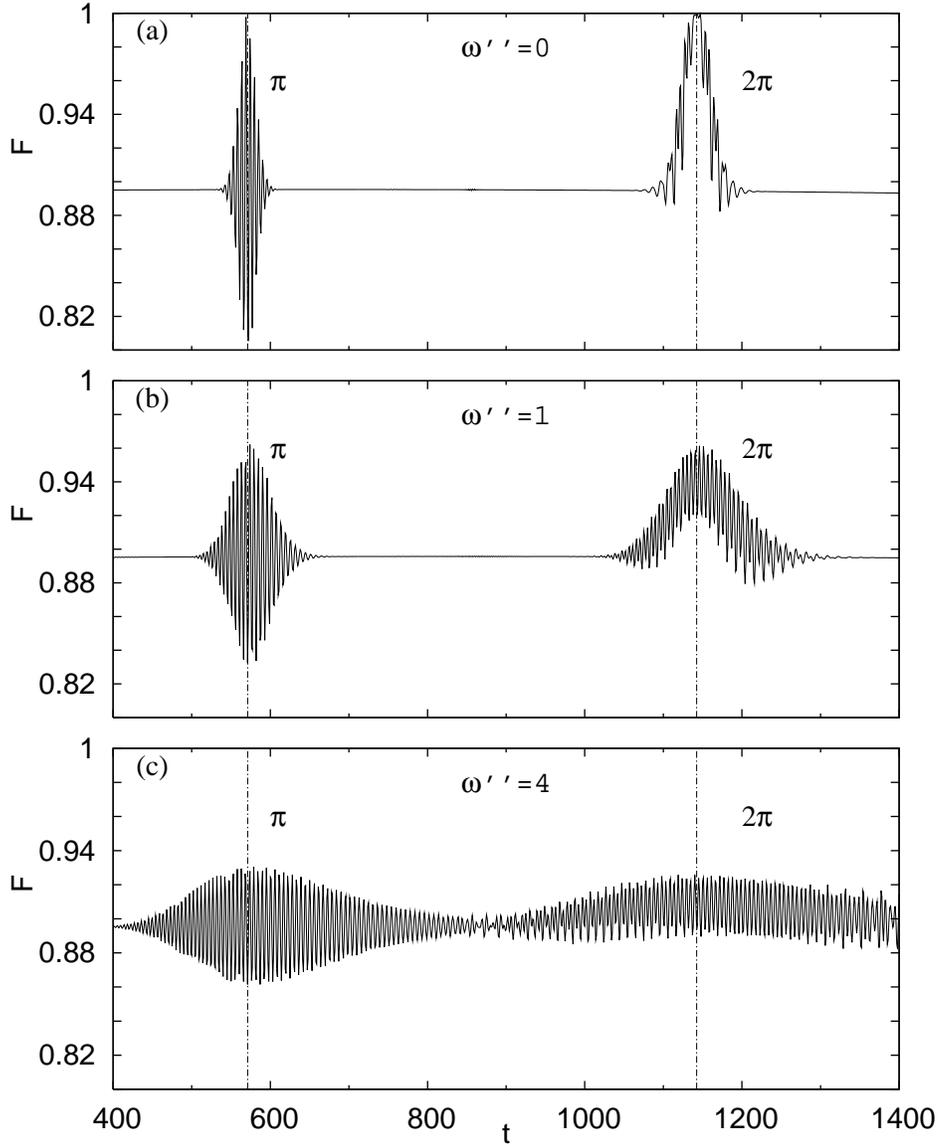}}
\caption{Structure of echo resonances for coherent initial states of the modified quantum top 
$U_0=\exp{(-\ii S [\alpha (S_z/S)^2/2 + \gamma (S_z/S - j^*)^3/6])}$, 
$\alpha=1.1, S=200, \delta=1.6 \cdot 10^{-3}$, for increasing value of $\omega''=\gamma=0$ (a),
$\omega''=1$ (b), and $\omega''=4$ (c), which weakens and broadens the resonances.
Note that in (a), $\omega''=0$ we have the same data as 
in figure \ref{fig:shortCoh}a. 
Vertical chain lines show theoretical times $t_{\rm r}/2$ for $\pi$, and 
$t_{\rm r}$ for $2\pi$  resonance.}
\label{fig:resonances}
\end{figure}   
Let us now discuss the behaviour of the fidelity for initial wavepackets 
in the regime of linear response approximation in some more detail. 
We shall consider possible deviations from the random phase approximation in the
time dependent exponentials of eqs. (\ref{eq:EVS},\ref{eq:EVS2}) which have been invoked previously in
order to derive time independent terms (\ref{eq:EVSA},\ref{eq:EVSA2}) of the fidelity plateau
(\ref{eq:plateauLR}), and also (\ref{eq:plateau}). Specifically we will explain the resonances observed 
{\em e.g.} in figure \ref{fig:shortCoh}.

For such a resonance to occur the phases of (\ref{eq:EVS},\ref{eq:EVS2}) have to build up in a
constructive way and this is clearly impossible in a generic case, unless: (i) We have one 
dimension $d=1$ so we sum up over a one-dimensional array of integers $n$ in the action space\footnote[2]{
In more than one dimension we would clearly need a strong condition on 
commensurability of frequency derivatives over the entire region of the action lattice 
where the initial state is supported.}.
(ii) The wavepacket is localized over a classically small region of the action 
space/lattice such that a variation of the frequency derivative 
$\d\omega(j)/\d j$ over this region is sufficiently small. 
The quantitative conditions for the occurrence together with the strength and the shape of 
such resonances are discussed below.

In this subsection we thus consider a one-dimensional case, $d=1$.  
Again we study time dependent terms of (\ref{eq:EVS},\ref{eq:EVS2}) which can all be cast into a 
general form (\ref{eq:gauss}), however, now the time is not small enough
to enable the sum over quantum number to be estimated by an integral. 
On the contrary, we seek a condition that the consecutive phases in the 
exponential build up an interference pattern.

\subsubsection{$2\pi$-resonance:}

Let us expand the frequency around the center of the packet
\begin{equation}
\omega(j) = \omega^* + (j-j^*)\omega' + \frac{1}{2}(j-j^*)^2\omega'' + \ldots
\end{equation}
where
\begin{equation}
\omega^* = \omega(j^*),\qquad \omega' = \frac{\d\omega(j^*)}{\d j},\qquad
\omega'' = \frac{\d^2\omega(j^*)}{\d j^2}.\qquad
\label{eq:omegaprime}
\end{equation}
The phases in the sums of the form (\ref{eq:gauss}) come into resonance, 
for $m = 1$ and hence for any higher $m\ge 1$, when they change by $2\pi$ per
quantum number, which happens at time $t_{\rm r}$ 
\begin{equation}
\hbar\omega' t_{\rm r} = 2\pi,\quad t_{\rm r} = \frac{2\pi}{\hbar\omega'}.
\label{eq:resonance}
\end{equation}
and its integer multiples\footnote[7]{It is interesting to note that
these resonant times correspond precisely to the condition for 
{\em revivals} of the wavepacket in the {\em forward} evolution (appart 
from a phase-space translation) studied in Refs.\cite{revivals}.}.
Let time $t$ be {\em close} to $k t_{\rm r}, k\in\Z$, and write $t = k t_{\rm r} + t'$ where 
$t' \ll t_{\rm r}$, so that 
\begin{equation}
\hbar \omega' t' \ll 2\pi.
\label{eq:approx}
\end{equation}
Now we can estimate the general time dependent term (\ref{eq:gauss}),
where $g(j)$ is again a smooth classical $\hbar$-independent function of $j$ representing
a suitable combination of Fourier coefficients $v_m(j)$, by: (i) shifting the time variable
to $t'$, (ii) incorporating the resonance condition (\ref{eq:resonance}), and (iii) 
approximating the resulting sum by an integral, due to smallness of $t'$ (\ref{eq:approx}), which is a simple
Gaussian:
\begin{eqnarray}
&& \sum_n g(\hbar n) e^{\ii m \omega(\hbar n)t} D_\psi(\hbar n)\nonumber\\ 
&\cong& \sum_n g(\hbar n)e^{\ii m \left(\omega^* + (\hbar n - j^*)\omega' + 
\frac{1}{2}(\hbar n -j^*)^2\omega''\right)(k t_{\rm r} + t')}D_\psi(\hbar n) 
\nonumber \\
&=& e^{\ii m (\omega^* t-\chi)}\sum_n g(\hbar n) e^{\ii m \left((\hbar n - j^*)\omega' t' + \frac{1}{2}(\hbar n - j^*)^2\omega'' t\right)} D_\psi(\hbar n) \nonumber \\
&\cong& 
e^{\ii m (\omega^* t-\chi)} g(j^*) \sqrt{\frac{\Lambda}{\pi\hbar}} \int\!\d j\,e^{\ii m \left((j-j^*)\omega' t' + \frac{1}{2}(j-j^*)^2\omega'' t\right) - \frac{\Lambda}{\hbar}(j-j^*)^2} \nonumber \\
&=& 
e^{\ii m (\omega^* t-\chi)} g(j^*) 
\left(1-\ii \frac{\hbar m\omega'' t}{2\Lambda}\right)^{-1/2}
\exp\left(-\frac{\hbar m^2\omega'^2 t'^2}{4\Lambda} \frac{1 + \ii\frac{\hbar m \omega'' t}{2\Lambda}}{1 + \left(\frac{\hbar m\omega'' t}{2\Lambda}\right)^2}\right)
\label{eq:resatom}
\end{eqnarray}
where $\chi=2\pi k j^*/\hbar$. From this calculation we deduce quantitative condition for the appearance and the shape of 
the resonance. Let $\Delta_j = \ave{(J-j^*)^2}^{1/2} = \sqrt{\frac{\hbar}{2\Lambda}}$ denote the 
action width of the wavepacket. Physically, we need that the coherence of linearly increasing 
phases is not lost along the size of the wavepacket, {\em i.e.}
\begin{equation}
\zeta := m\omega'' t \Delta_j^2 = \frac{\hbar m \omega'' t}{2\Lambda} < 2\pi.
\label{eq:ResCond}
\end{equation}
$\zeta$ is precisely the coefficient appearing in the square-root prefactor and the exponential 
of the resonance profile (\ref{eq:resatom}).
Indeed we see that with increasing $\zeta$ the squared modulus of the peak of the resonance 
(at $t'=0$) dies out as $(1 + \zeta^2)^{-1/2}$. We note that $\zeta$ increases with the increasing
order $k$ of the resonance, since $t \approx k t_{\rm r}$, 
so 
\begin{equation}
\zeta = k m \frac{\pi}{\Lambda} \frac{\omega''}{\omega'}.
\label{eq:ResCond1}
\end{equation}
Therefore we may get strong and numerous resonances, {\em i.e.} small $\zeta$,
provided either the second derivative $\omega''$ is small, or the initial state is squeezed 
such that $\Lambda \gg 1$.
For example, if the second derivative vanishes everywhere, $\omega''\equiv 0$, 
then the resonances may appear even for extended states. This is the case for our numerical model, where resonances can be seen also for a random state in figure \ref{fig:shortRan}.

From (\ref{eq:resatom}) we read that the temporal profile in such a fidelity resonance 
has a shape of a 
Gaussian of effective width  
\begin{equation}
\Delta_t = \frac{\sqrt{1 + \zeta^2}}{m \omega' \Delta_j},
\end{equation}
modulated with an oscillation of frequency $\cong \omega^*$. 
Hence in order to feel the effect of the fidelity
resonance, time $t$ has to be within $\Delta_t$ of the center of the resonance $k t_{\rm r}$. 
In the semiclassical limit, the resonance positions scale as 
$t_{\rm r}\propto \hbar^{-1}$, while their widths grow only as $\Delta_t\propto \hbar^{-1/2}$, so they are well separated.
Also, with increasing order $k$ the magnitudes of the peaks of the resonances decrease as
$\sim k^{-1/2}$, while their widths increase as $\sim k$, so they will eventually, at 
$k \propto \hbar^{-1/2}$, start to overlap. This will happen at time $\sim \hbar^{-3/2}$ which 
is smaller than $\sim t_2$ provided $\delta < \nu^{-1/2}\hbar$.

The resonance described in this paragraph, which will be called a $2\pi$-resonance, 
affects {\em all} time dependent terms of (\ref{eq:EVS},\ref{eq:EVS2}), 
but its precise shape depends on coefficients $\tilde{v}_m(j^*)e^{\ii m\theta^*}$.
However, it is important to note that the fidelity can be explicitly calculated
close to the center of the resonance, $t' \ll \Delta_t$, where the Gaussian factor of the right-most 
expression of (\ref{eq:resatom}) can be neglected. In addition we shall neglect the coefficient
$\zeta$ in (\ref{eq:resatom}) as we are particularly interested in the case of a strong 
resonance $\zeta \ll 1$. Then a simple calculation gives for the first moment
of $\Sigma_t$ (\ref{eq:EVS})
\begin{equation}
\ave{\Sigma_t} \cong \sum_{m\neq 0} \tilde{v}_m(j^*) e^{\ii m \theta^*}\left(1 - e^{\ii m \omega^* t}\right),
\end{equation}
while the second moment can be shown to be just $\ave{\Sigma^2_t} = \ave{\Sigma_t}^2$,
hence the fidelity is, around the center of a $2\pi$-resonance, equal to $1$ 
within the linear response approximation
\begin{equation}
F(t) \cong 1,\qquad {\rm for}\quad |t-k t_{\rm r}| \ll \Delta_t,\;\; {\rm and}\;\; \zeta \ll 1.
\end{equation}

We note that such a `flat-top' structure of $2\pi-$resonance is nicely illustrated in a numerical example in 
figure \ref{fig:resonances}, where we consider a slightly modified model with
$U_0=\exp{(-\ii S [\alpha (S_z/S)^2/2 + \gamma (S_z/S - j^*)^3/6])}$,
and $h_0(j) = \alpha j^2/2 + \gamma (j-j^*)^3/6$, such that $\omega'' = \gamma $ may not be
identically vanishing.
 
\subsubsection{$\pi$- and $2\pi/m$-resonances:}
\begin{figure}[h]
\vspace{-1cm}
\centerline{\includegraphics[angle=-90,width=140mm]{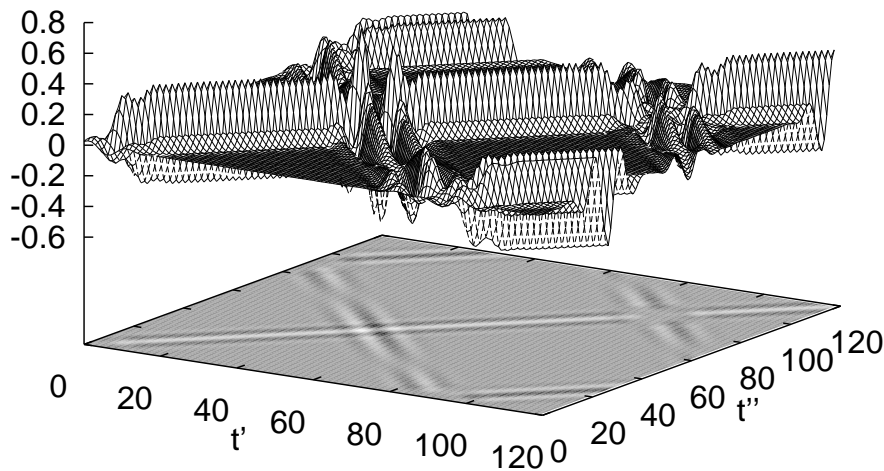}}
\vspace{-1cm}
\caption{The two-time correlation function $C(t',t'')$ (\ref{eq:Ct}) for the quantized top with 
$\alpha=1.1, \beta=0, \gamma=0, S=16$, and coherent initial state.}
\label{fig:Cor}
\end{figure}   
We note that one may obtain a resonance condition for time dependent term (\ref{eq:resatom})
with fixed $m$ for even shorter time, namely for $t = t_{\rm r}/m$. This is trivially the
case for perturbations with many, or at least more than one Fourier components with $|m|> 1$.
However, in such cases only selected time dependent terms of the moments (\ref{eq:EVS},\ref{eq:EVS2})
will be affected, so the fidelity will generically {\em not come back to} $1$, even in the linear 
response regime (\ref{eq:LR2}) and in the strongly resonant case $\zeta \ll 1$. 
Such (incomplete) resonances at fractional times 
$(k/m)t_{\rm r}$ will be called $2\pi/m$-resonances.

However, we may obtain a resonant condition at $t=t_{\rm r}/2$ even for the first Fourier 
component $m=1$ of the perturbation, due to taking the square of the operator 
$\Sigma_t$ thus producing Fourier number $m+m'=\pm 2$ in the last term on the RHS of eq. 
(\ref{eq:EVS2}). Such a resonant behaviour at times $(k + \frac{1}{2})t_{\rm r}$ will 
be called a $\pi-$resonance.

So for perturbations with a {\em single} Fourier mode $m=\pm 1$, or more generally with only
odd-numbered Fourier modes $m=2l+1$, the $\pi-$resonance can affect only the last term of 
the second moment (\ref{eq:EVS2})
while it cannot affect the first moment (\ref{eq:EVS}), which is given by its time averaged value (\ref{eq:EVSCS}). To see this, we observe that the time dependent parts of form $g(\hbar n) e^{\ii m \omega t}$ in $\ave{\Sigma_t}$ and $\ave{\Sigma_t^2}$ are proportional to $\sum_n{D_\psi(\hbar n) g(\hbar n) (-1)^{mn}}$. As $m$ is an odd number and $D_\psi(\hbar n) g(\hbar n)$ is a smooth function of $n$ this sum averages to zero. All this allows us to again explicitly 
compute the linear response fidelity (\ref{eq:LR2}) close to the peak in a strongly resonant case, 
namely
\begin{eqnarray}
\ave{\Sigma_t} \cong \sum_{m=2l+1} \tilde{v}_m(j^*) e^{\ii m \theta^*},\quad{\rm for}\quad 
|t-(k+\frac{1}{2})t_{\rm r}| \ll 
\Delta_t,\;\;{\rm and}\;\;\zeta \ll 1,\\
\ave{\Sigma_t^2} \cong \Bigl(\sum_{m=2l+1} \tilde{v}_m(j^*) e^{\ii m \theta^*}\Bigr)^2 
+ \Bigl( \sum_{m=2l+1} \tilde{v}_m(j^*) e^{\ii m (\theta^*+\omega^* t)}\Bigr)^2, \\
F(t) \cong 1 - \frac{4\delta^2}{\hbar^2}\left(\sum_{m=2l+1}^{m>0} |\tilde{v}_m(j^*)|\cos(m\omega^* t + \beta_m)\right)^2,
\end{eqnarray}
where $\beta_m$ are phases of complex numbers $\tilde{v}_m(j^*)e^{\ii m\theta^*}$.
So we have learned that the fidelity at the peak of a $\pi$-resonance displays 
an oscillatory pattern, oscillating precisely around the plateau value 
$F_{\rm plateau}$ (\ref{eq:LRCS}) with an amplitude of oscillations equal to 
$1-F_{\rm plateau}$ so that the fidelity comes back to $1$ close to the peak
of the resonance.

Again, our numerical example illustrates such an oscillatory structure of $\pi-$resonance in 
figure \ref{fig:resonances}. The resonances can also be nicely seen in `short-time' figure 
\ref{fig:shortCoh}, and because $\zeta=0$ also in the 'long-time' figure \ref{fig:longCoh}. 
%But note that there we do not plot every point for $S=1600$ figures 
%(time axis is rescaled with $\sqrt{S}$) and so the structure of resonances can not be seen exactly.
In figure \ref{fig:Cor} we depict the structure of $\pi-$ and $2\pi-$ resonance as reflected
in the two-time correlation function $C(t',t'')$. Note that the first intersection of the
soliton-like-trains for $t'-t''={\rm const}$ and $t+t'={\rm const}$ happens at $t_{\rm r}/2$ and
produces a $\pi-$resonance, while the second intersection at $t_{\rm r}$ produces a $2\pi$-resonance.

In analogy to the emergence of a $\pi-$resonance as a consequence of the contribution from the 
{\em second} moment of $\Sigma_t$, even for the first Fourier mode $m=1$, we shall eventually 
obtain also fractional $2\pi/p$-resonances
at times $(k/p)t_{\rm r}$ in the non-linear-response regimes where higher moments 
$\ave{\Sigma^p_t}$ contribute to $F(t) \sim 
\ave{\exp(\ii\Sigma_t \delta/\hbar)}$, eq. (\ref{eq:plateau}).
This is illustrated numerically in figure \ref{fig:shortCoh}c showing the case of strong 
perturbation $\delta S=3.2$ so that higher orders are important. One indeed obtains 
fractional resonances, some of which have been marked on the figure.

\subsection{Illustration in terms of echoed Wigner functions}
\label{subsect:wig}

\begin{figure}[h]
\centerline{\includegraphics[angle=0,width=110mm]{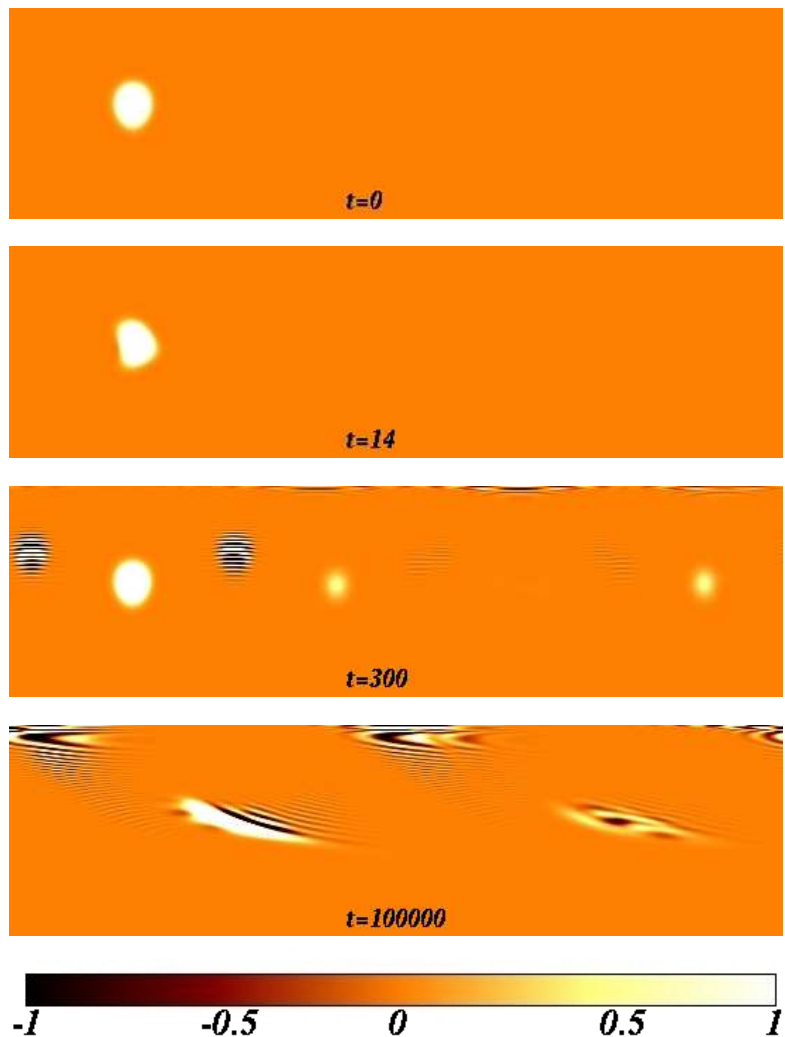}}
\caption{(Movies online \cite{mov})
Snapshots of the Wigner function of echo-dynamics for quantized top 
$\alpha=1.1,\beta=0,\gamma=0$ with $S=200$ and for $\delta=1.6\cdot 10^{-3}$ 
(same as in figures \ref{fig:shortCoh}a,\ref{fig:resonances}a).
The upper phase-hemi-sphere is shown with $j=\cos\vartheta\in[0,1]$ on the
vertical axis and $\theta=\varphi\in [0,2\pi]$ on the horizontal axis.
From top to bottom we show: the initial state at $t=0$, the state at 
$t=14\approx t_1$ when we are around the regular Ehrenfest time, 
at $t=300$ in the 
middle of the plateau, and at $t=100000$ in the ballistic regime. The bar below
shows the color-code of the Wigner function values.}
\label{fig:movies}
\end{figure}

All the phenomena described theoretically in the preceeding subsections
can be nicely illustrated in terms of the {\em echoed Wigner function} --- 
the Wigner function $W_{M_\delta(t)\ket{\psi}}(\vec{q},\vec{p})$ 
of the echo-dynamics. Namely, 
according to formula (\ref{eq:Wigecho}) the fidelity $F(t)$ is given
simply by the overlap of the echoed Wigner function and the
Wigner function of the initial state.
Therefore, the phase-space chart of the echoed Wigner function
contains the most detailed information on echo-dynamics 
and illustrates the essential differences between different regimes of
fidelity decay.
This is shown in figure~\ref{fig:movies} (see \cite{mov})
for the quantized top where the Wigner function on a sphere is 
computed according to Ref.\cite{agarwal}.
In the initial {\em classical regime}, $t < t_1$, the echoed Wigner function
has not yet developed negative values and is in point-wise agreement with 
the Liouville density of the classical echo-dynamics.
In the {\em plateau regime}, $t_1 < t < t_2$, the echoed Wigner function
decomposes into several pieces, one of which freezes at the position of the 
initial packet providing significat and constant overlap --- the plateau.
At very particular values of time, namely at the echo resonances, 
different pieces of the echoed Wigner function somehow magically recombine
back into the initial state (provided $\zeta \ll 1$).
In the asymptotic, {\em ballistic regime}, $t > t_2$, even the frozen piece
starts to drift ballistically away from the postion of the initial packet,
thus explaining a fast Gaussian decay of fidelity.

\section{Semiclassical asymptotics: random initial state}
\label{sec:SARS}

The second specific case of interest is that of a random initial state.
Here we shall assume that our Hilbert space has a finite dimension 
${\cal N}$, like e.g. in the case of the kicked top or a general quantum map
with a finite classical phase space, or it is determined by some
large classically invariant region of phase space, e.g. we may consider
all states $\ket{\vec{n}}$ between two energy
surfaces $E_1 < h_0(\hbar\vec{n}) < E_2$ of an autonomous system. 
In any case we have the scaling
\begin{equation}
{\cal N}\cong \frac{{\cal V}}{\hbar^d}
\end{equation}
where ${\cal V}$ is the classical $d-$volume of the populated action space 
region of interest.
The notion of a random state refers to an ensemble average over the
full Hilbert space of interest. So we treat the complex coefficients
$\psi_{\vec{n}}$ as pairs of components of a vector on a $2{\cal N}$-dimensional
unit sphere. In the asymptotic regime of large ${\cal N}$, these can in turn be
replaced by independent complex Gaussian variables with the variance
\begin{equation}
\aave{\psi^*_{\vec{n}'}\psi_{\vec{n}}} \cong 
\frac{1}{\cal N}\delta_{\vec{n}'\vec{n}},\quad
\aave{\psi_{\vec{n}'}\psi_{\vec{n}}} \cong 0,
\label{eq:cf}
\end{equation}
where $\aave{\bullet}$ denotes an ensemble average over random states.
When we write such an average for an operator, we actually mean
\begin{equation}
\aave{A}:=\aave{\bra{\psi}A\ket{\psi}} = \frac{1}{\cal N}\tr A.
\end{equation}
Note that a trivial application of the pair-contraction rule (Wick theorem) yields that averaged 
fidelity is asymptotically the same as the averaged fidelity amplitude squared \cite{Maribor}
\begin{equation}
\aave{F(t)} = \aave{|f(t)|^2} = |\aave{f(t)}|^2 + \frac{1}{\cal N} \cong
|\aave{f(t)}|^2.
\end{equation}
This means that the fidelity amplitude is self-averaging, {\em i.e.} its variance with respect to random
state averaging is semiclassically vanishing. The same property holds for the fidelity itself,
$\aave{F^2} - \aave{F}^2 = {\cal O}(1/{\cal N})$.

\subsection{The plateau: Linear response and beyond}

\begin{figure}[h!]
\centerline{\includegraphics{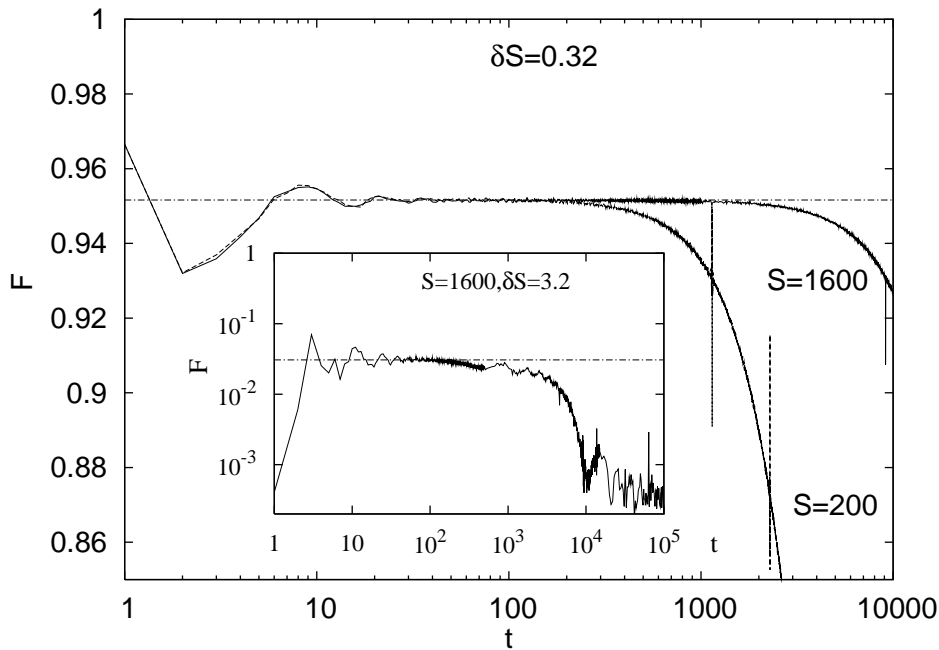}}
\caption{Short time fidelity for a quantized top with $\alpha=1.1, \beta=1.4$ and random initial state. 
To reduce statistical fluctuations, averaging over $20$ realizations of initial random states is 
performed for $S=1600$, and over $100$ initial states for $S=200$. The horizontal chain line is the
semiclassical theory (\ref{eq:ResCond}). Resonances are here present due to the special
property $\omega''(j)=0$ and will be absent for a more generic unperturbed system. The main figure shows the case of weak perturbation $\delta S = 0.32$,
whereas the inset shows the case of strong perturbation $\delta S = 3.2$.} 
\label{fig:shortRan}
\end{figure}   

\begin{figure}[h!]
\centerline{\includegraphics{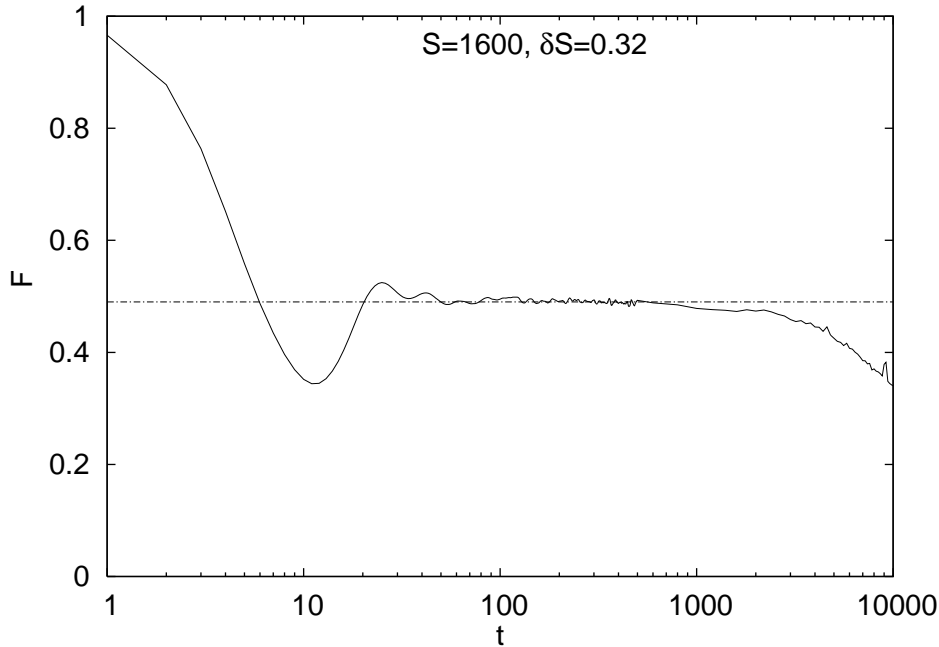}}
\caption{Short time fidelity for $\alpha=1.1, \beta=0, S=1600, \delta S=0.32$ and random initial state. The chain line shows the theoretical value of the plateau as computed from the formula (\ref{eq:plateauRS1}).
}
\label{fig:shortRanSing}
\end{figure}   

In computing the ensemble average of the linear response formula
(\ref{eq:LR2}) we have to compute ensemble averages of the expressions
(\ref{eq:EVS},\ref{eq:EVS2}). This is a straightforward application
of (\ref{eq:cf}) for (\ref{eq:EVS2}), 
and the pair-contraction rule for (\ref{eq:EVS})
\begin{eqnarray}
\aave{\Sigma_t^2} &\cong& \frac{\tau^2}{\cal N}
\sum_{\vec{n}}\sum_{\vec{m}\neq 0} |v_{\vec{m}}(\hbar\vec{n})|^2
\frac{\sin^2\left(\frac{1}{2}\vec{m}\cdot\vec{\omega}(\hbar\vec{n})t\right)}
{\sin^2\left(\frac{1}{2}\vec{m}\cdot\vec{\omega}(\hbar\vec{n})\right)} \nonumber\\
&\cong& \frac{\tau^2}{\cal V}\int\d^d\vec{j}
\sum_{\vec{m}\ne\vec{0}}|v_{\vec{m}}(\vec{j})|^2
\frac{\sin^2\left(\frac{1}{2}\vec{m}\cdot\vec{\omega}(\vec{j})t\right)}
{\sin^2\left(\frac{1}{2}\vec{m}\cdot\vec{\omega}(\vec{j})\right)},
\label{eq:EVSR2} \\
\aave{\ave{\Sigma_t}^2} &\cong& \frac{1}{{\cal N}}\aave{\Sigma^2_t}.
\label{eq:EVSR}
\end{eqnarray}
We note that the expression (\ref{eq:EVSR2}) is just the classical phase space
average $\ave{\sigma_t^2}_{\rm cl}$ where $\sigma_t(\vec{j},\vec{\theta}) = 
\sum_{t'=0}^{t-1} v(\vec{j},\vec{\theta} + \vec{\omega}(\vec{j})t')$ is the
classical limit of $\Sigma_t$.
We see that for a random state the contribution of the square
of the expectation value (\ref{eq:EVSR}) is semiclassically small so the
plateau in fidelity is within the linear response approximation determined 
by the second moment (\ref{eq:EVSR2}).

On one hand, let us assume that
\begin{equation}
\vec{m}\cdot\vec{\omega}(\vec{j}) \neq 0 \pmod{2\pi}
\label{eq:nonres}
\end{equation}
for all contributing Fourier components $\vec{m}$ and
for all $\vec{j}$ from the classical action space of interest.
In such a case the $\sin^2$ in the denominator of 
(\ref{eq:EVSR2}) is never vanishing, so the integral has no problem
with singularities, while the $\sin^2$ in the numerator can be
averaged, either over a short period of time, or over 
small regions of phase space for sufficiently long time $t>t_1$, in both cases
yielding a trivial factor of $1/2$. In any case, the convergence timescale $t_1$
here is classical, and of course, does not depend on the
strength of the perturbation $t_1 \sim \hbar^0 \delta^{0}$.
Hence in such a {\em non-singular} case, the linear response 
plateau of the fidelity reads
\begin{eqnarray}
&&\aave{F(t)} \cong 1 - 
\frac{\delta^2}{\hbar^2}\nu_{\rm ran} + {\cal O}(\delta^4),\\
&&\nu_{\rm ran} = \frac{\tau^2}{\cal V}\int\d^d\vec{j}\sum_{\vec{m}\neq\vec{0}}
\frac{|v_{\vec{m}}(\vec{j})|^2}
{2\sin^2(\frac{1}{2}\vec{m}\cdot\vec{\omega}(\vec{j}))}.
\quad
\label{eq:LRRS}
\end{eqnarray}
Going beyond the linear response approximation we again use the general 
formula (\ref{eq:plateau}), and compute the expectation value of an operator in the random state in 
terms of the classical phase-space average 
\begin{equation}
\aave{g(\vec{J},\vec{\Theta})}\cong \frac{1}{(2\pi)^d{\cal V}}
\int\d^d\vec{j}\int\d^d\vec{\theta} g(\vec{j},\vec{\theta}),
\end{equation}
that is
\begin{equation}
\aave{f_{\rm plateau}} \cong \frac{1}{\cal V}\int\d^d\vec{j}\left|
\int\frac{\d^d\vec{\theta}}{(2\pi)^d} 
\exp\left(\frac{i\delta}{\hbar}\tilde{v}(\vec{j},\vec{\theta})\right)\right|^2.
\label{eq:plateauRS}
\end{equation}

If, on the other hand, the condition (\ref{eq:nonres}) does not hold, i.e.
if there exist values of the action $\vec{j}$, and $\vec{m}\in\Z^d$, $k\in\Z$, 
such that $\vec{m}\cdot\vec{\omega}(\vec{j}) = 2\pi k$,
then the denominator of (\ref{eq:EVSR2}) becomes {\em singular} and the corresponding term
(in the appropriate limit) grows in time.
However this growth stops, at least on a timescale $\sim\hbar^{-1}$, 
due to a discrete nature of the quantum action space.
At this time the value of the correlation integral, the plateau, will
be typically so small that it cannot be described within the linear response
approximation, so we have to employ the
general formula (\ref{eq:plateau}).
The only modification of the general formula, with respect to a non-singular
case (\ref{eq:plateauRS}), is the observation that the quantum phase space is
in fact discrete in action so one should semiclassically approximate the
expectation value with the sum, instead of an integral,
\begin{equation}
\aave{g(\vec{J},\vec{\Theta})} \cong \frac{1}{\cal N} \sum_{\vec n}
\frac{1}{(2\pi)^d}\int\d^d \vec{\theta}
g(\hbar\vec{n},\vec{\theta}).
\end{equation}
Furthermore, the diverging terms 
$v_{\vec{m}}/\sin{(\vec{m}\cdot \vec{\omega}/2)}$ in $\tilde{v}_\vec{m}$ 
(\ref{eq:tilde}) come from the semiclassical approximation for 
$\Sigma_t$ whose quantum counterpart has matrix elements proportional to 
$V_{mn}/\sin{[(\phi_m-\phi_n)/2]}$. As we consider perturbations with a zero average, 
$V_{mm}\equiv 0$, the diverging terms are absent in the quantum operator $\Sigma_t$. 
Therefore to remedy the random state formula for $f_{\rm plateau}$ (\ref{eq:plateauRS}) we simply 
replace an integral with the summation excluding the divergent terms,
so the general final result reads
\begin{equation}
\aave{f_{\rm plateau}} \cong \frac{1}{\cal N}\sum_{\vec{n}}^{\vec{m}\cdot\vec{\omega}(\hbar\vec{n})\neq
2\pi k}\left|
\int\frac{\d^d\vec{\theta}}{(2\pi)^d} 
\exp\left(\frac{\ii\delta}{\hbar}\tilde{v}(\hbar\vec{n},\vec{\theta})\right)\right|^2.
\label{eq:plateauRS1}
\end{equation}

Again we find an excellent confirmation of our theoretical predictions in the numerical
experiment. In the first calculations we choose the shift $\beta = 1.4$ so that we have no
singular frequency throughout the action space. 
In figure \ref{fig:shortRan} we demonstrate the plateau, which in the case of random states starts 
earlier than for coherent states, namely at $t_1 \propto \hbar^0 \delta^0$. The value of the plateau 
can be calculated by numerically evaluating the integrals occurring in eq.  (\ref{eq:LRRS}), for the
linear response approximation, or eq. (\ref{eq:plateauRS}) in general. The integral over the 
angle $\theta$ in the formula for the plateau (\ref{eq:plateauRS}) again gives a Bessel function so that we end up with a numerical integration over $j$
\begin{equation}
\aave{F_{\rm plateau}} \cong \left[ \frac{1}{2}\int_{-1}^1\d j J_0^2\left(\delta S\frac{\sqrt{1-j^2}}{2\sin{\left\{ \alpha (j-\beta)/2\right\}}}\right) \right]^2.
\label{eq:BessR}
\end{equation}
Observe that the random state plateau is just a square of the action-space average of a 
coherent state expression (\ref{eq:Plat}). Horizontal chain lines in figure \ref{fig:shortRan} correspond to this theoretical values and agree with 
the numerics, both for weak perturbation $\delta S = 0.32$ and strong perturbation $\delta S = 3.2$ (inset). The plateau lasts up to $t_2$ which is for random states $\hbar$-independent, 
$t_2 \sim 1/\delta$. Small resonances visible in the figures are due to the fact that the 
Hamiltonian is a quadratic function of the action and therefore $\omega'' \equiv 0$, so that the 
resonance condition (\ref{eq:resonance}) is satisfied also for extended states (\ref{eq:ResCond}). 
For a more generic Hamiltonian these narrow resonant spikes will be absent. 
In figure.~\ref{fig:shortRanSing} we also demonstrate the plateau in the fidelity for the
zero-shift case $\beta=0$ with a singular-frequency, $\omega(j=0) = 0$, where we again find an excellent
agreement with the theoretical prediction (\ref{eq:plateauRS1}). In this case the theoretical value has been 
obtained by replacing an integral in (\ref{eq:BessR}) with a sum over $n$ (replacing $j= \hbar n$) and summing over all quantum numbers except $n=0$. Observe that the value of the plateau is much lower than in the case of 
a non-zero shift $\beta=1.4$ in figure \ref{fig:shortRan}.

\subsection{Asymptotic regime of long times}

\begin{figure}[h]
\centerline{\includegraphics{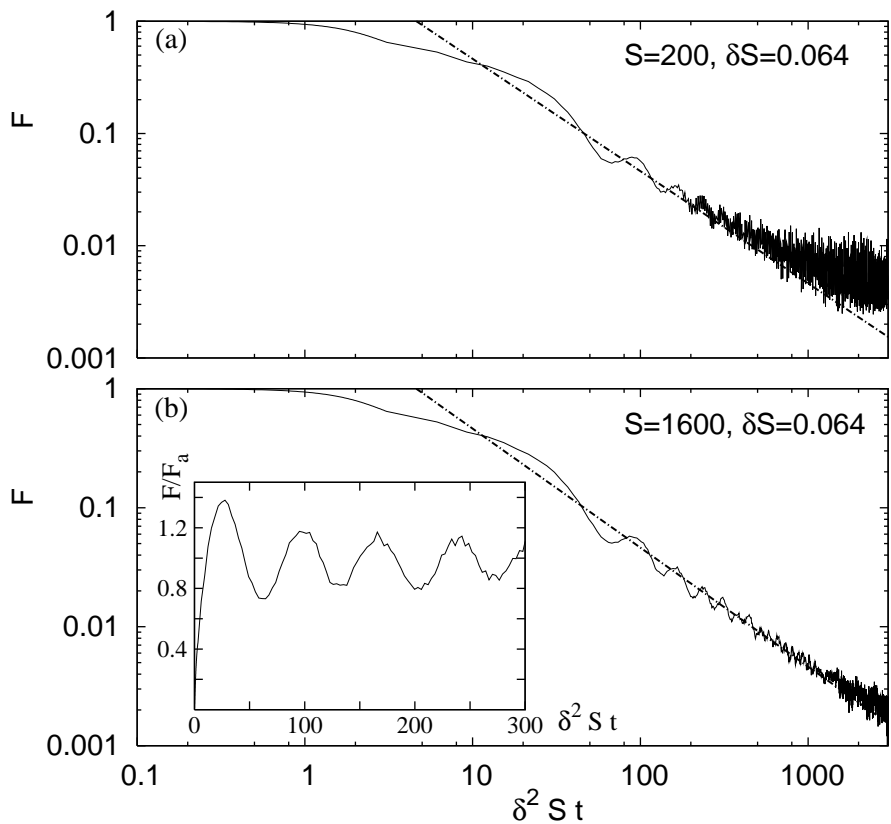}}
\caption{Long time fidelity for random states for quantized top with $\alpha=1.1,\beta=0$, for
$S=200$ (a), and $S=1600$ (b). 
Here $\delta S=0.064$ and averaging for $S=200$, and $S=1600$, is performed over $1000$, and $20$, initial random states, respectively. Heavy chain line is the theoretical asymptotic decay (\ref{eq:fergr}) with analytically computed prefactor
(no fitting parameters). The inset in the bottom figure shows the diffractive quotient between the
numerical fidelity and the asymptotic formula (\ref{eq:Fsqrt}) (chain line in the main figure).}
\label{fig:longRan}
\end{figure}
Again, after sufficiently long time $t_2 \sim \delta^{-1}$,
the second term in BCH expansion (\ref{eq:BCH1}) will be dominating
and we shall use the ensemble average of the ASI formula (\ref{eq:ferg2})
where $\aave{D_\psi} = \frac{1}{\cal N}$
\begin{equation}
\aave{f(t)} \cong \frac{1}{\cal V} \int \d^d\vec{j} 
\exp\left(\ii\frac{\tau\delta^2}{2\hbar}\bbar{v}(\vec{j})t\right).
\label{eq:fergr}
\end{equation}
The semiclassical computation of such an integral is an
elementary application of stationary phase method in $d$-dimensions,
following \cite{PZ} for the analogous case of a generic observable.
The condition for the validity of stationary phase method is
that $t\delta^2/\hbar > 1$ which will turn out to be consistent with
the assumption $t > t_2$. Let $\vec{j}_\eta, \eta=1,\ldots,p$ be the
$p$ points where the phase of the exponential on RHS of (\ref{eq:fergr})
is stationary, 
$\partial\bbar{v}(\vec{j}_\eta)/\partial\vec{j}=0$. 
This yields a simple result
\begin{equation}
\aave{f(t)} \cong
\frac{(2\pi)^{3d/2}}{\cal V}\left|\frac{\hbar}{t\tau\delta^2}\right|^{d/2}
\sum_{\eta=1}^p\frac{
\exp\left\{\ii\bbar{v}(\vec{j}_\eta)t\tau\delta^2/(2\hbar) + i \nu_\eta \right\}}{|\det W_\eta|^{1/2}},
\label{eq:Fsqrt}
\end{equation}
where 
\begin{equation}
(W_\eta)_{jk} := \frac{\partial^2 \bbar{v}(\vec{j}_\eta)}
{\partial j_k\partial j_l}
\end{equation}
is a matrix of second derivatives at the stationary point $\eta$, and 
$\nu_\eta = \pi(m_+ - m_-)/4$
where $m_{\pm}$ are numbers of positive/negative eigenvalues of the matrix
$W_\eta$. Here we should remember that the asymptotic formula (\ref{eq:fergr}) has been obtained as
a stationary phase approximation of an integral in the limit of an infinite action space. 
If we have a finite region of the action space, the stationary phase 
approximation of (\ref{eq:fergr}) gives additional oscillating 
prefactor, whose amplitude dies out as $(\hbar/t)^{1/2}$ for $\hbar\to 0$ and/or $t\to\infty$, and which can be interpreted as a {\em diffraction}. 
This oscillating prefactor can be seen in numerical data for fidelity in the inset of 
figure \ref{fig:longRan}.

So we have found that, apart from possible oscillation due to phase 
differences if $p>1$, the fidelity will for a random state asymptotically decrease with a power law
\begin{equation}
\aave{F(t)} \sim \left(\frac{t}{t_{\rm ran}}\right)^{-d},
\quad t_{\rm ran} = {\rm const} \frac{\hbar}{\delta^2}.
\end{equation}
Note that for a random initial state, the actual transition time scale $t_2$, as predicted by
eq.(\ref{eq:t2}), indeed does not depend on $\hbar$.
 
The above theory is again quantitatively confirmed in figure \ref{fig:longRan} for
the quantized top with $\alpha=1.1,\beta=0$,
where a (single) stationary point needed for the formula (\ref{eq:Fsqrt}) had to be calculated
numerically.

\section{Discussion}

In the present paper we have elaborated a semiclassical theory of quantum fidelity decay 
for systems with an integrable classical counterparts, perturbed by observables of vanishing time
average. Such perturbations may not be generic, but provide important special class of 
perturbations which are often enforced by symmetries.

We have found that quantum fidelity will generally, after initial decay on a short perturbation
independent timescale $t_1$, exhibit a saturation around a constant value --- the plateau, and stay there up to time $t_2$, such that the time span of the plateau $t_2/t_1 \sim 1/\delta$ can be made
arbitrary long for small perturbation $\delta$. After the plateau, $t > t_2$, the fidelity will decay
as a Gaussian for a coherent initial state, or as a power law $t^{-d}$ for random initial states,
just to name the two most important specific cases, where the timescale of decay is generally
proportional to $\delta^{-2}$. This must be contrasted with the decay in the regular 
case of a non-zero time-average perturbation \cite{PZ} where the decay time scales with the 
perturbation strength as $\delta^{-1}$.
In the case of localized initial wavepackets in one dimension, we observe and explain the
effect of echo resonances, {\em i.e.} the sudden revivals of fidelity at perturbation independent
and equally spaced instants of time.

The freezing of fidelity is a distinct quantum phenomenon, as the corresponding classical 
fidelity for the initial Gaussian wavepacket displays a power-law decay $t^{-d}$ \cite{Veble} after the 
point $t_1$ where the quantum plateau starts. 
The classical fidelity decays on a time scale $\delta^{-1}$ no matter what the average value of the perturbation is, while the time scale of quantum fidelity decay drastically changes from $\delta^{-1}$ to $\delta^{-2}$, having $\bar{V}=0$. This increased stability of regular quantum systems to 
perturbations with a zero time average could be potentially useful in constructing quantum 
devices \cite{qcomp}. This is even more so because the plateau also exists for random initial states
which are expected to be more relevant for efficient quantum information processing.

\section*{Acknowledgements}

Useful discussions with T. H. Seligman, G. Veble, G. Benenti and G. Casati are gratefully 
acknowledged. The work has been financially supported by the Ministry of Science, Education 
and Sport of Slovenia, and in part by the grant DAAD19-02-1-0086, ARO, United States.


\section*{References}

\begin{thebibliography}{1}

\bibitem{peres}
Peres A 1984 Phys.~Rev.~A {\bf 30} 1610--15; see also 
Peres A 1995 {\it Quantum Theory: Concepts and Methods} (Dordrecht: Kluwer).

\bibitem{jalabert}
Jalabert R A and Pastawski H M 2001 Phys.~Rev.~Lett.~ {\bf 86} 2490

\bibitem{generalrefs} 
Pastawski H M {\em et al.} 1995 Phys.~Rev.~Lett. {\bf 75} 4310;\\
Levstein P R {\em et al.} 1998 J.~Chem.~Phys. {\bf 108} 2718;\\
Cerruti N R and Tomsovic S 2002 Phys.~Rev.~Lett. {\bf 88} 054103;\\
Cucchietti F M {\em et al.} 2002 Phys.~Rev.~E {\bf 65} 046209;\\
Wisniacki D A {\em et al.} 2002 Phys.~Rev.~E {\bf 65} 055206;\\
Wisniacki D A and Cohen D 2002 Phys.~Rev.~E {\bf 66} 046209;\\
Berman G P {\em et al} 2002 Phys.~Rev.~E {\bf 66} 056206;\\
Wang W G and Baowen Li 2002 Phys.~Rev.~E {\bf 66} 056208;\\
Benenti G and Casati G 2002 Phys.~Rev.~E {\bf 66} 066205;\\
Jacquod Ph {\em et al.} 2002, Phys.~Rev.~Lett. {\bf 89} 154103;\\
Karkuszewski Z P {\em et al.} 2002, Phys.~Rev.~Lett. {\bf 89} 170405;\\
Weinstein Y S {\em et al.} 2002 Phys.~Rev.~Lett. {\bf 89} 214101;\\
Emerson J {\em et al} 2002 Phys.~Rev.~Lett. {\bf 89} 284102;\\
Kottos T and Cohen D 2003 Europhys. Lett. {\bf 61} 431;\\
Wisniacki D A 2003 Phys.~Rev.~E {\bf 67} 016205;\\
Silvestrov P G {\em et al} 2003 Phys.~Rev.~E. {\bf 67} 025204;\\
M.~F.~Andersen, A.~Kaplan, N.~Davidson 2003, Phys.~Rev.~Lett. {\bf 90} 023001;\\
\v Znidari\v c M and Prosen T 2003, J.~Phys.~A:Math. Gen. {\bf 36} 2463;\\
Benenti G, Casati G and Veble G 2003, Phys.~Rev.~E {\bf 67} 055202(R);\\
Vanicek J and Heller E J 2003, preprint {\tt quant-ph/0302192};\\
Sankaranarayanan R and Lakshminarayan A 2003, preprint {\tt nlin.CD/0307005};\\
Goussev A and Dorfman J R 2003, preprint {\tt nlin.CD/0307025}.

\bibitem{Prosen01} Prosen T, 2002 Phys.~Rev.~E {\bf 65} 036208

\bibitem{QC} Prosen T and \v Znidari\v c M 2001 J.~Phys.~A: Math. Gen. {\bf 34} L681

\bibitem{PZ} Prosen T and \v Znidari\v c M 2002 J.~Phys.~A: Math. Gen. {\bf 35} 1455

\bibitem{golden} Jacquod Ph, Silvestrov P G and Beenakker C W J 
2001 Phys.~Rev.~E {\bf 64} 055203(R)

\bibitem{Veble} Benenti G, Casati G and Veble G 2003, preprint {\tt nlin.SI/0304032}

\bibitem{Bruno} Eckhardt B 2003 J.~Phys.~A: Math. Gen {\bf 36} 371

\bibitem{wrongpaper} Jacquod Ph, Adagideli I and Beenakker C W J 2003 Europhys. Lett. {\bf 61} 729

\bibitem{Fox} Fox R F and Elston T C 1994 Phys.~Rev.~E {\bf 50} 2553

\bibitem{Maribor} Prosen T, Seligman T H and  \v Znidari\v c M 2003, preprint {\tt quant-ph/0304104}

\bibitem{EBK} See e.g. Berry M V 1977 Philos. Trans. R. Soc. London A {\bf 287} 237

\bibitem{revivals} Braun P A and Savichev V I 1996 J. Phys. B: At. Mol. Opt. Phys. {\bf 29} L329; see also related general discussion in Leichtle C, Averbukh I Sh and Schleich W P 1996 Phys.~Rev.~A {\bf 54} 5299

\bibitem{mov} See movies at 
\href{http://chaos.fiz.uni-lj.si/papers/freeze}{\tt http://chaos.fiz.uni-lj.si/papers/freeze}. 

\bibitem{agarwal} Agarwal G S 1981 Phys.~Rev.~A {\bf 24} 2889

\bibitem{qcomp} Nielsen M A and Chuang I L 2000 {\em Quantum computation and quantum information} (Cambridge University Press)

\end{thebibliography}
\end{document}